%% file: main.tex
\documentclass[final,3p,times,twocolumn,authoryear,square,comma,sort&compress]{elsarticle}
\pdfoutput=1
\usepackage{amssymb}
\usepackage{subfig}
\usepackage{caption}
\usepackage{xspace}
\usepackage{longtable}
\usepackage{pdflscape}
\usepackage{multirow}
\usepackage{ragged2e}
\usepackage{booktabs}
\usepackage{slashed}
\usepackage{hyperref}
\usepackage{amsmath}
\usepackage{cuted}
\usepackage[switch]{lineno}
\usepackage{booktabs}
\usepackage{lipsum}
\usepackage{natbib}
\setcitestyle{numbers,square}
\journal{Physical Letter B}

\DeclareUnicodeCharacter{2212}{-}
\begin{document}
\begin{frontmatter}
\author[USTC]{Xinbai. Li}
\author[USTC]{Jiaxuan. Luo}
\author[USTC]{Zebo. Tang}
\author[USTC]{Xin. Wu}
\author[USTC]{Wangmei. Zha\corref{cor1}}
\ead{first@ustc.edu.cn}
\affiliation[USTC]{organization={University of Science and Technology of China},
    city={Hefei},
    postcode={230026},
    country={China}}
\title{Exploring the higher-order QED effects on the differential distributions of Breit-Wheeler process in relativistic heavy-ion collisions}



%
\begin{abstract}

Extensive studies have been conducted in the past few decades to investigate potential signatures of higher-order QED effects in high-energy electromagnetic scattering processes. In our previous work, we have identified evidence of higher-order corrections in the total cross-section for the Breit-Wheeler process in relativistic heavy-ion collisions. However, the presence of higher-order QED corrections cannot be unambiguously proven solely based on total cross-section measurements due to substantial experimental and theoretical uncertainties. The objective of this paper is to explore the sensitivity of specific differential observables in the Breit-Wheeler process to higher-order QED effects in high-energy heavy-ion collisions. These investigations will provide guidance in determining the presence or absence of higher-order QED processes by conducting precise measurements in future experiments.

\end{abstract}


\begin{keyword}
QED \sep Pair production \sep Higher Order \sep Heavy Ion Collisions 
\end{keyword}

\end{frontmatter}

\section {Introduction}
\label{Introduction}

\input{Introduction}
\input{Theoretical1}
\input{results}

\input{summary1}
\input{acknowledge}

\bibliographystyle{elsarticle-num} 
\bibliography{main}

\end{document}

%% file: Introduction.tex
In 1934, Breit and Wheeler proposed the theory of the reverse process of the Dirac annihilation to create electron-positron pair from the collision of two photons, known as the Breit-Wheeler process~\cite{PhysRev.46.1087}. While they acknowledged the experimental challenge of observing this process directly, they suggested that the strong electromagnetic field of highly charged nuclei could serve as a suitable photon source~\cite{WW, PhysRev.45.729}. After almost 90 years, the STAR collaboration at RHIC experimentally verified this process in heavy-ion collisions, providing definitive confirmation of Breit and Wheeler's original proposal~\cite{STAR:2019wlg}.

In heavy-ion collisions involving large Z nuclei, the coupling constant Z$\alpha$ ($\sim$ 0.6 for gold and lead) in the Breit-Wheeler process approaches 1, indicating the presence of significant higher-order effects. In 1954, Bethe, Maximon, and Davies conducted a pioneering study on higher-order effects in quantum electrodynamics (QED)~\cite{PhysRev.93.768, PhysRev.93.788}. They investigated these effects in a similar process known as the Bethe-Heitler process~\cite{BH}, which involves the photoproduction of electron-positron pairs in the nuclear Coulomb field. They introduced the Sommerfeld-Maue approach, which incorporates a negative correction proportional to $Z^2$ in the total cross-section. In the case of relativistic heavy-ion collisions involving the Breit-Wheeler process, a stronger correction is expected due to the attachment of the quasi-real photon from the projectile to the heavy nuclei, in contrast to the Bethe-Heitler process~\cite{PhysRevD.57.4025, kk}.

The search for higher-order QED effects in heavy-ion collisions has been ongoing for several decades~\cite{RHIC1, RHIC2}, driven by the development of relativistic heavy-ion facilities such as RHIC at Brookhaven and LHC at CERN during the late 20th century. However, all measurements are found to be in excellent agreement with the calculations based on the equivalent photon approximation approach (EPA)~\cite{baltz2008physics,PhysRevC.97.054903,aaboud2018observation,adam2018low}, specifically employing the industry-standard model known as STARlight~\cite{starlight}. These calculations yield results identical to the lowest order QED predictions for cross-section estimation~\cite{Zha:2018tlq}. Some theorists argue that higher-order corrections to the Breit-Wheeler process at RHIC and LHC, in the relevant kinetic regions, are negligibly small~\cite{BAUR20071,PhysRevD.57.4025,SUN2020135679,BALTZ2001395}. This perspective is supported by an intuitive picture where, in the center-of-mass frame, a lepton pair within this kinematic regime behaves as a charge-neutral object in the Coulomb field. On the other hand, other theorists developed an appropriate regularization technique by introducing Coulomb potential screening~\cite{PhysRevA.61.032103, PhysRevA.64.032106}. This regularization leads to a significant negative correction in the measured cross-sections at RHIC and LHC. Within this framework, hints of higher-order effects were observed compared to the measurements at RHIC, which contradicts the conclusions drawn from the EPA calculations~\cite{PhysRevLett.100.062302, zha2021discovery}.

In our previous work, we found that traditional EPA calculations fail to account for the production of lepton pairs within the geometric radius of the nucleus~\cite{zha2021discovery}. Considering this factor, the lowest-order QED calculations overestimate the global cross-section measurements of the Breit-Wheeler process in heavy-ion collisions, thus providing evidence for the presence of higher-order effects. However, significant experimental uncertainties, including systematic uncertainty and uncertainties arising from luminosity determination~\cite{baltz2008physics,PhysRevC.97.054903,aaboud2018observation,adam2018low}, prevent the cross-section measurements alone from unambiguously proving the presence of higher-order corrections. In this paper, we utilize the theoretical framework established in our previous work and focus on examining particular differential observables that exhibit sensitivity to higher-order QED effects while being robust against systematic uncertainty and uncertainties associated with luminosity determination.

\begin{figure}[ht]
  \centering
  \subfloat[parallel diagram]{\includegraphics[width=0.43\columnwidth]{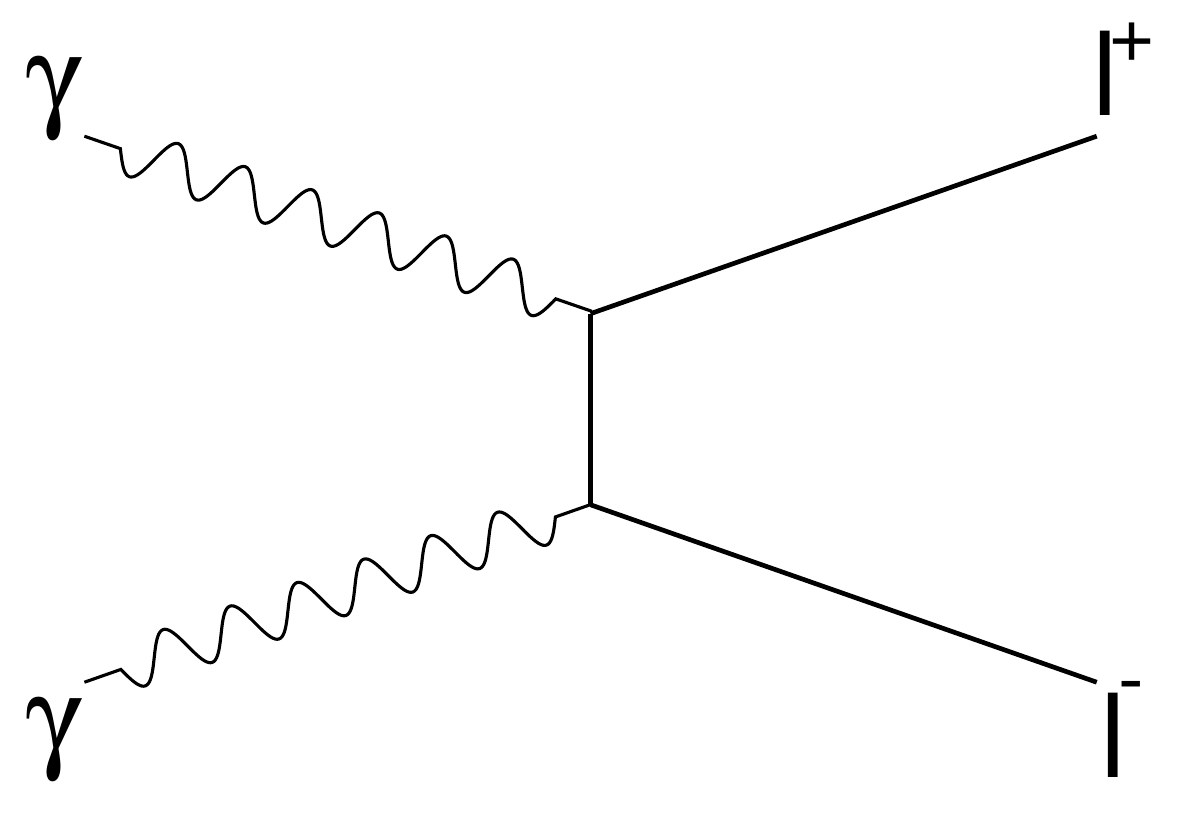}}\hspace{8mm}
  \subfloat[cross diagram]{\includegraphics[width=0.43\columnwidth]{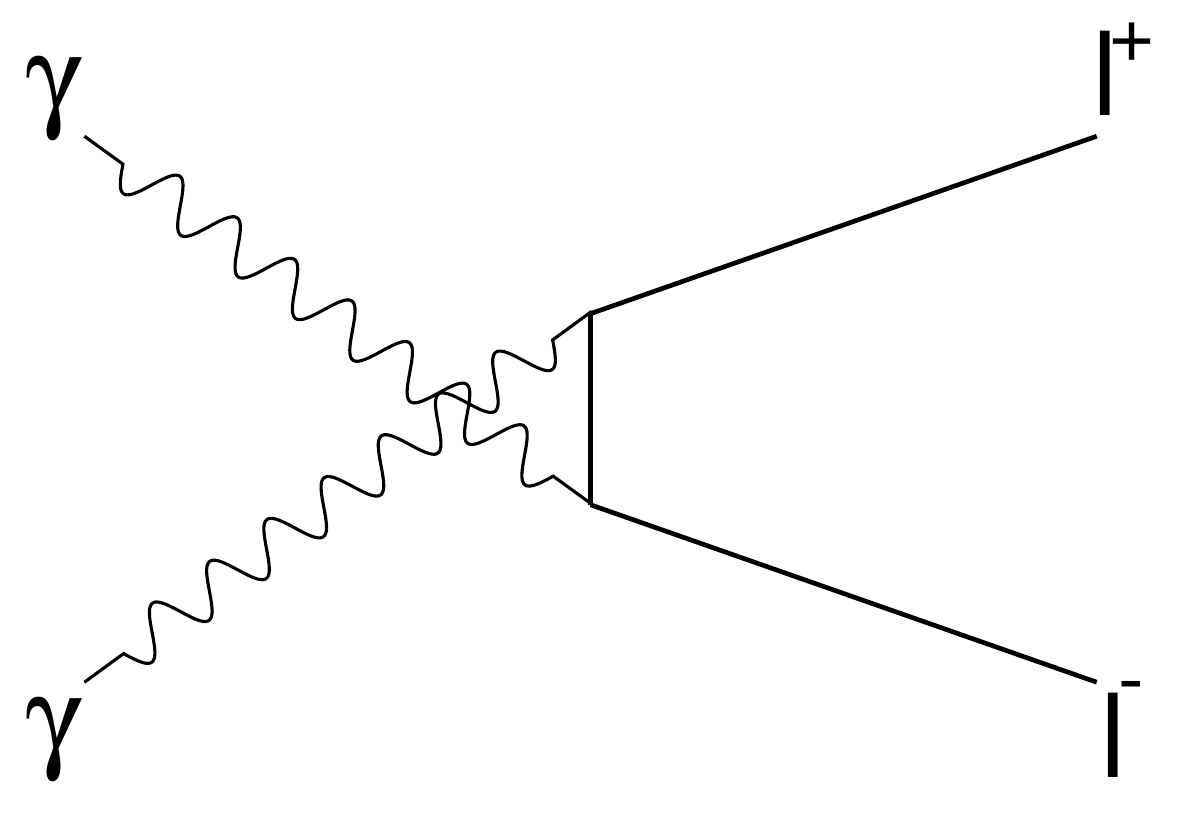}}
\caption{Two lowest order Feynman diagrams for lepton pair production from two photon fusion.}
\label{feynman}
\end{figure}

\begin{table*}
\caption{\textbf{Fiducial cuts implemented in the calculation}}
\centering
\begin{tabular}{llccccc}
\toprule
\multicolumn{2}{c}{Process and beam energy} & $p_{Tl}$ (GeV/c)& $\eta_{l}$ & $P_{Tll}$ (GeV/c) &$Y_{ll}$ & $M_{ll}$ (GeV)\\
\midrule
$\gamma \gamma \rightarrow e^+ e^-(\mu^+ \mu^-)$ &Au+Au $\sqrt{s_{NN}}=200 $ GeV & $(0.2,+\infty)$& $(-1.0,1.0)$& $(0,0.3)$& $(-1.0,1.0)$& $(0.4,2.6)$\\
$\gamma \gamma \rightarrow e^+ e^-$ &Pb+Pb $\sqrt{s_{NN}}=5.02 $ TeV& $(0.5,+\infty)$& $(-1.0,1.0)$& $(0,0.3)$&  $(-1.0,1.0)$ &$(1.0,2.8)$\\
$\gamma \gamma \rightarrow \mu^+ \mu^- (\tau^+ \tau^-)$ &Pb+Pb $\sqrt{s_{NN}}=5.02$ TeV & $(4.0,+\infty)$& $(-2.4,2.4)$& $(0,0.3)$& $(-2.4,2.4)$ & $(8.0,100.0)$\\
\bottomrule
\end{tabular}
\label{table}
\end{table*}
\begin{figure*}[ht]
  \centering
  \subfloat[$\gamma \gamma \rightarrow e^+ e^-$ in Au+Au $\sqrt{s_{NN}}=200 $ GeV]{\includegraphics[width=0.33\linewidth]{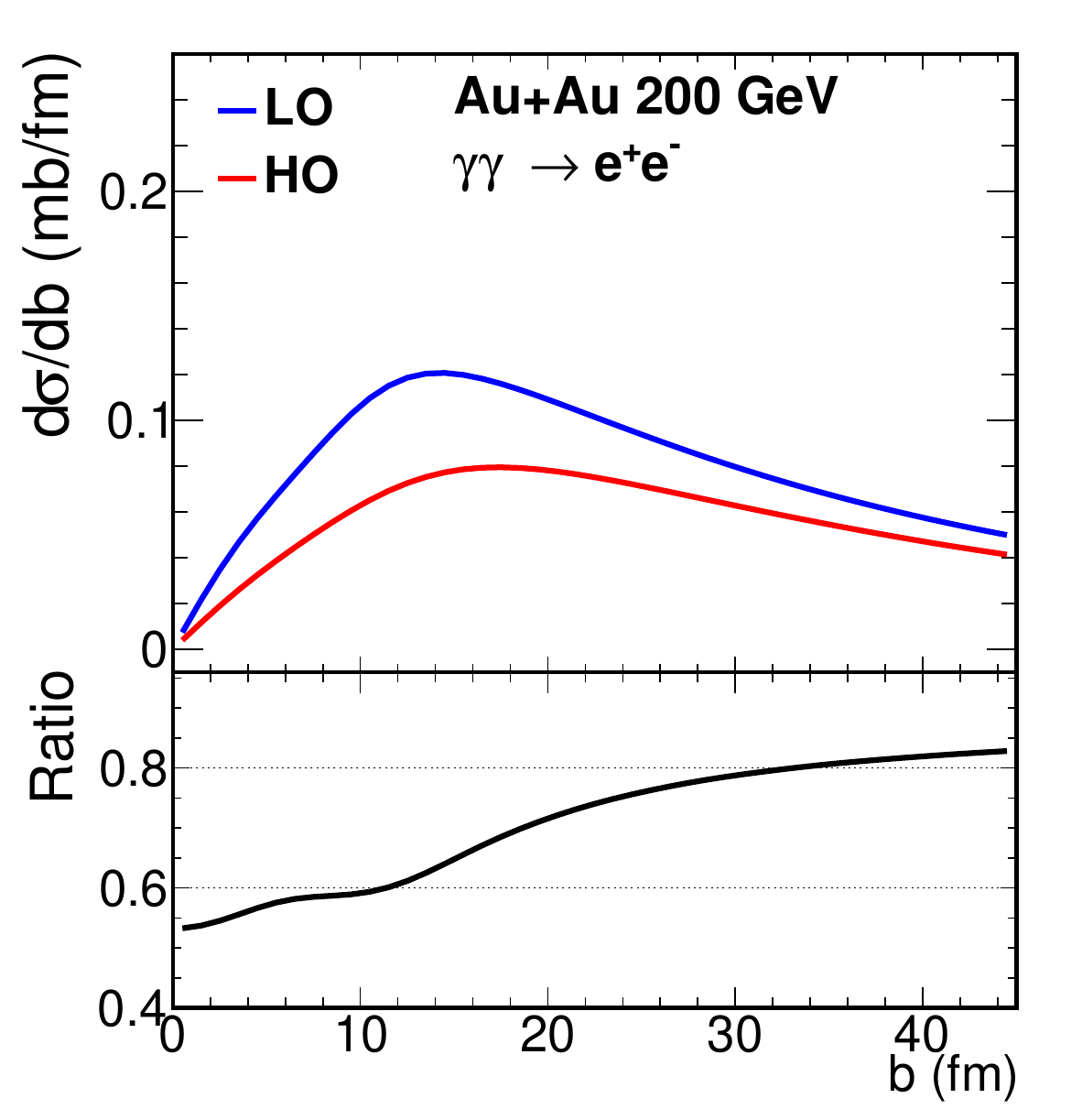}}
  \subfloat[$\gamma \gamma \rightarrow \mu^+ \mu^-$ in Au+Au $\sqrt{s_{NN}}=200 $ GeV]{\includegraphics[width=0.33\linewidth]{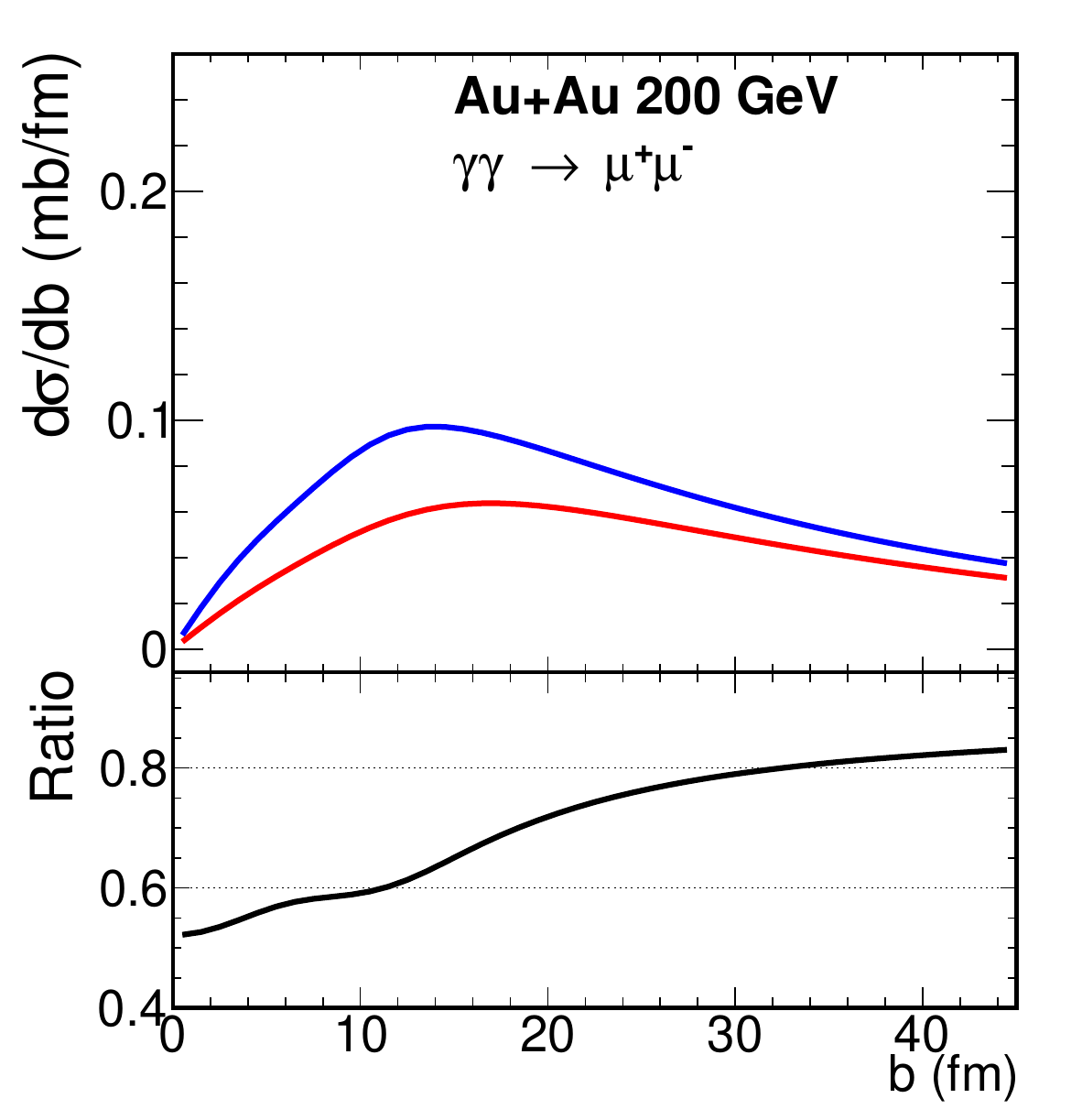}}
  \quad
  \subfloat[$\gamma \gamma \rightarrow e^+ e^-$ in Pb+Pb $\sqrt{s_{NN}}=5.02 $ TeV]{\includegraphics[width=0.33\linewidth]{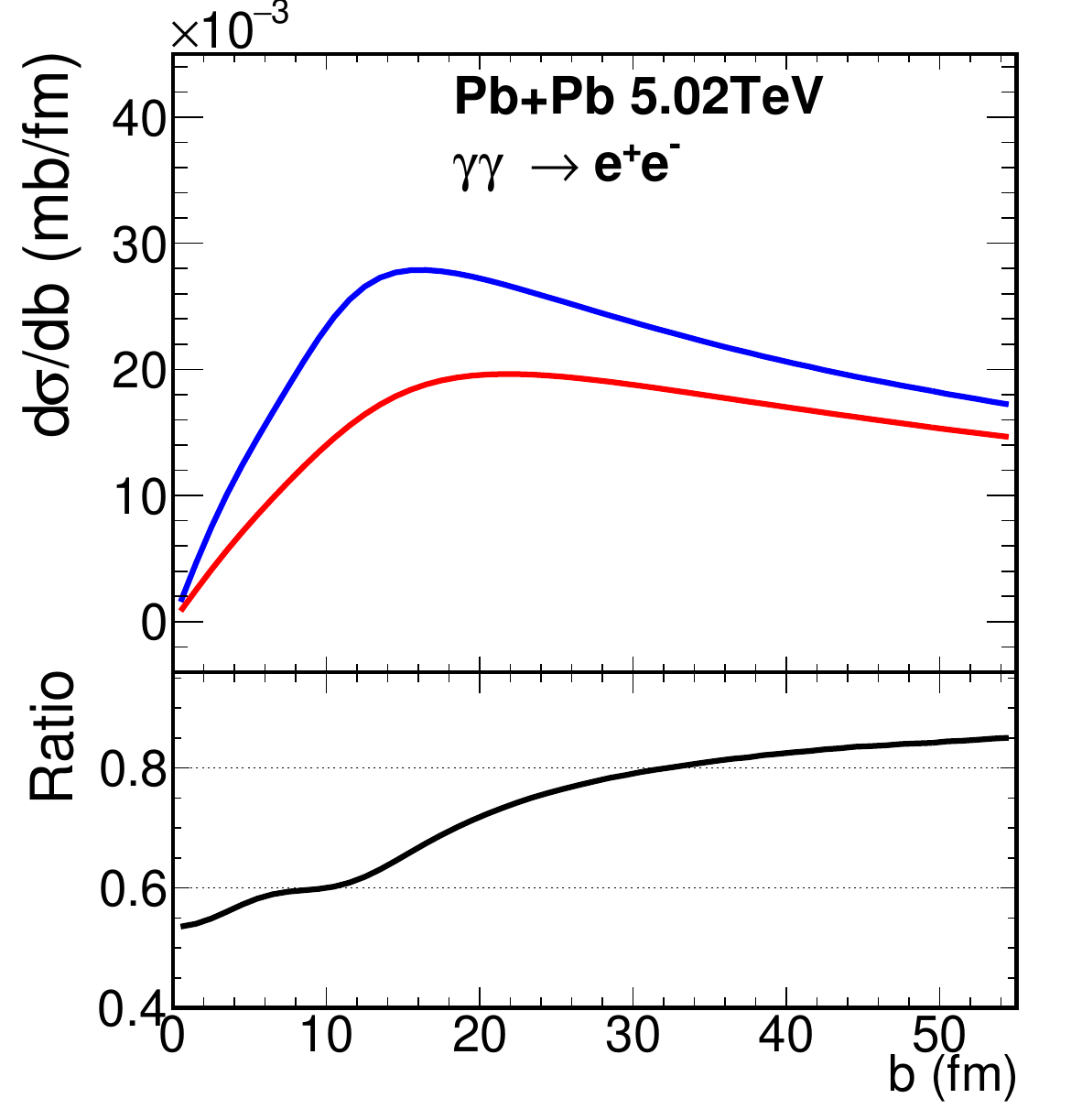}}
  \subfloat[$\gamma \gamma \rightarrow \mu^+ \mu^-$ in Pb+Pb $\sqrt{s_{NN}}=5.02 $ TeV]{\includegraphics[width=0.33\linewidth]{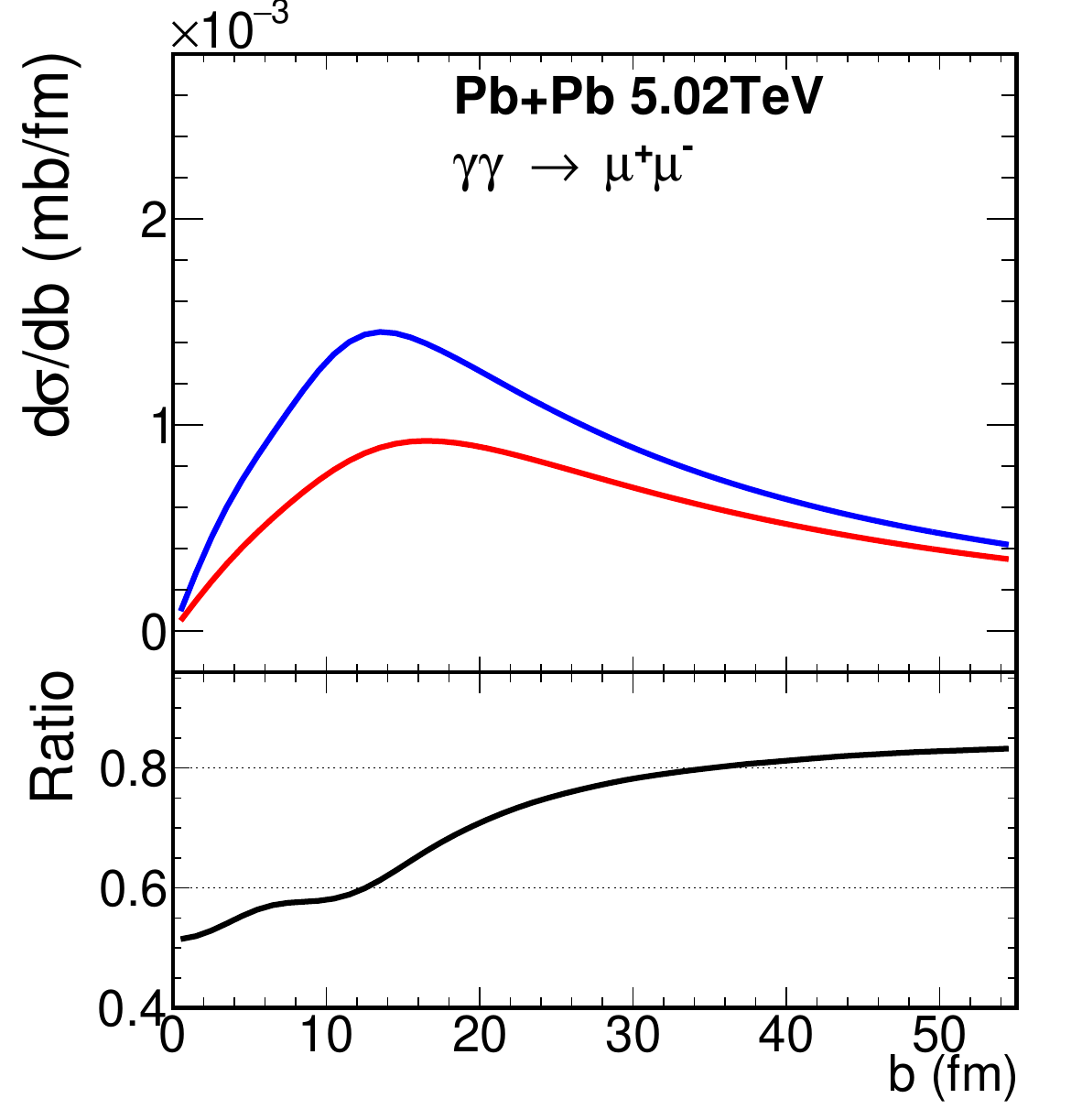}}
  \subfloat[$\gamma \gamma \rightarrow \tau^+ \tau^-$ in Pb+Pb $\sqrt{s_{NN}}=5.02 $ TeV]{\includegraphics[width=0.33\linewidth]{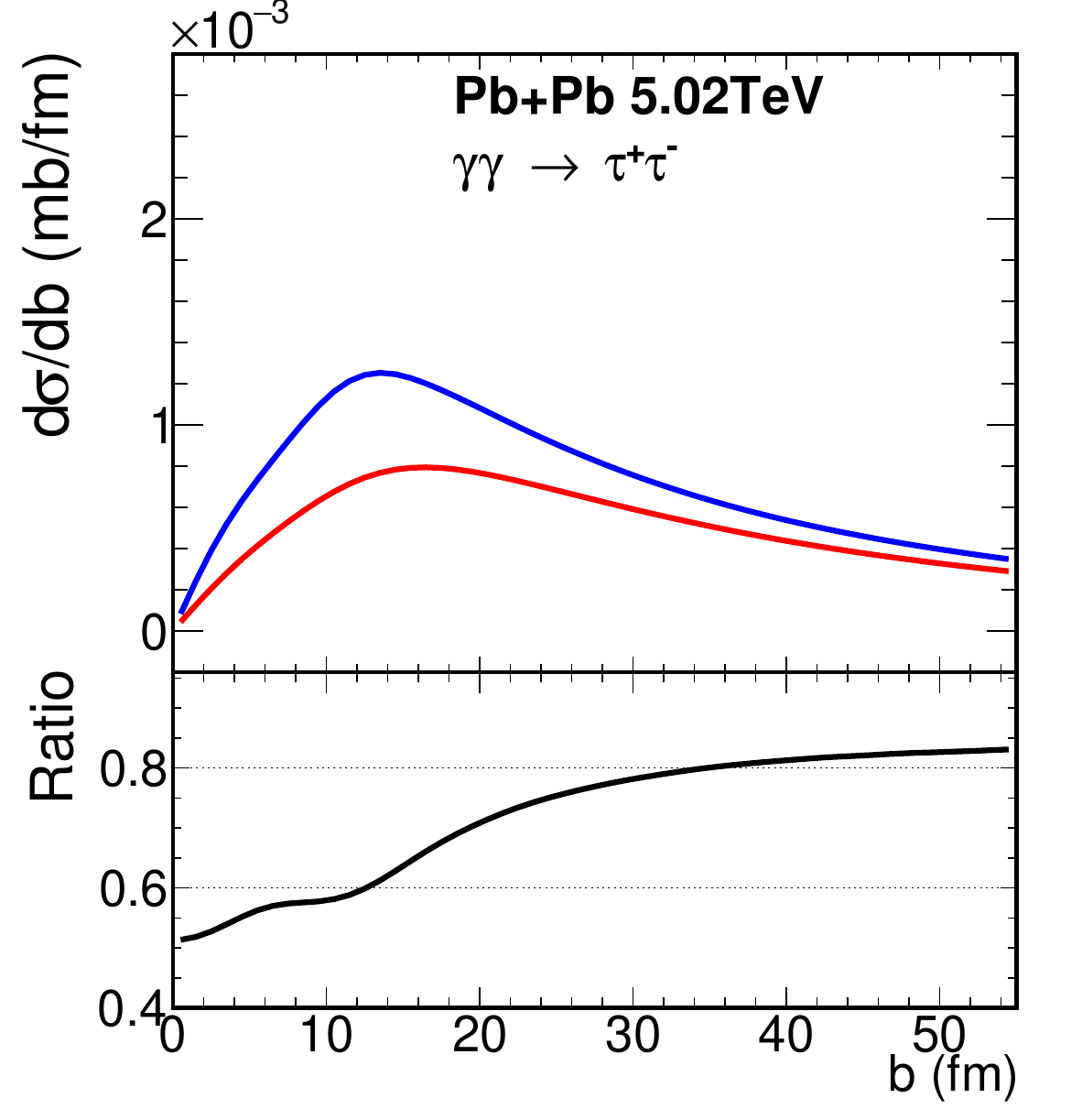}}
 \hfill
\caption{The cross-section of lepton pair production as a function of impact parameter from the lowest-order and higher-order calculations and the ratios of higher-order results (red lines) to the lowest-order (blue lines) results at typical RHIC and LHC collision energy and species. The fiducial cuts for the cross-section calculations are described in Tab.~\ref{table}. }
\label{yield}
\end{figure*}
\begin{figure*}[ht]
  
  \centering
  \subfloat[$\gamma \gamma \rightarrow e^+ e^-$ in Au+Au $\sqrt{s_{NN}}=200 $ GeV]{\includegraphics[width=0.33\linewidth]{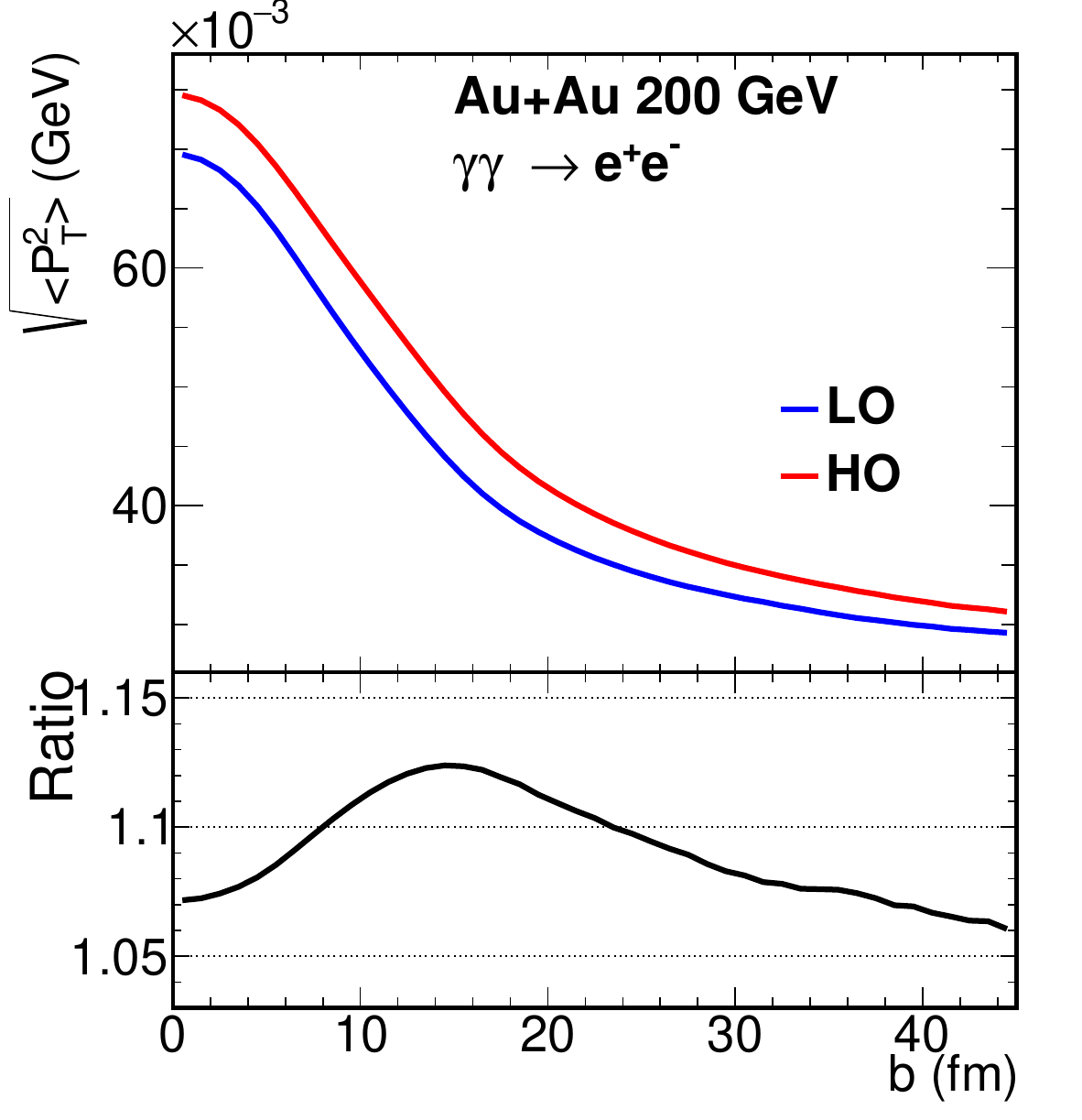}}
  \subfloat[$\gamma \gamma \rightarrow \mu^+ \mu^-$ in Au+Au $\sqrt{s_{NN}}=200 $ GeV]{\includegraphics[width=0.33\linewidth]{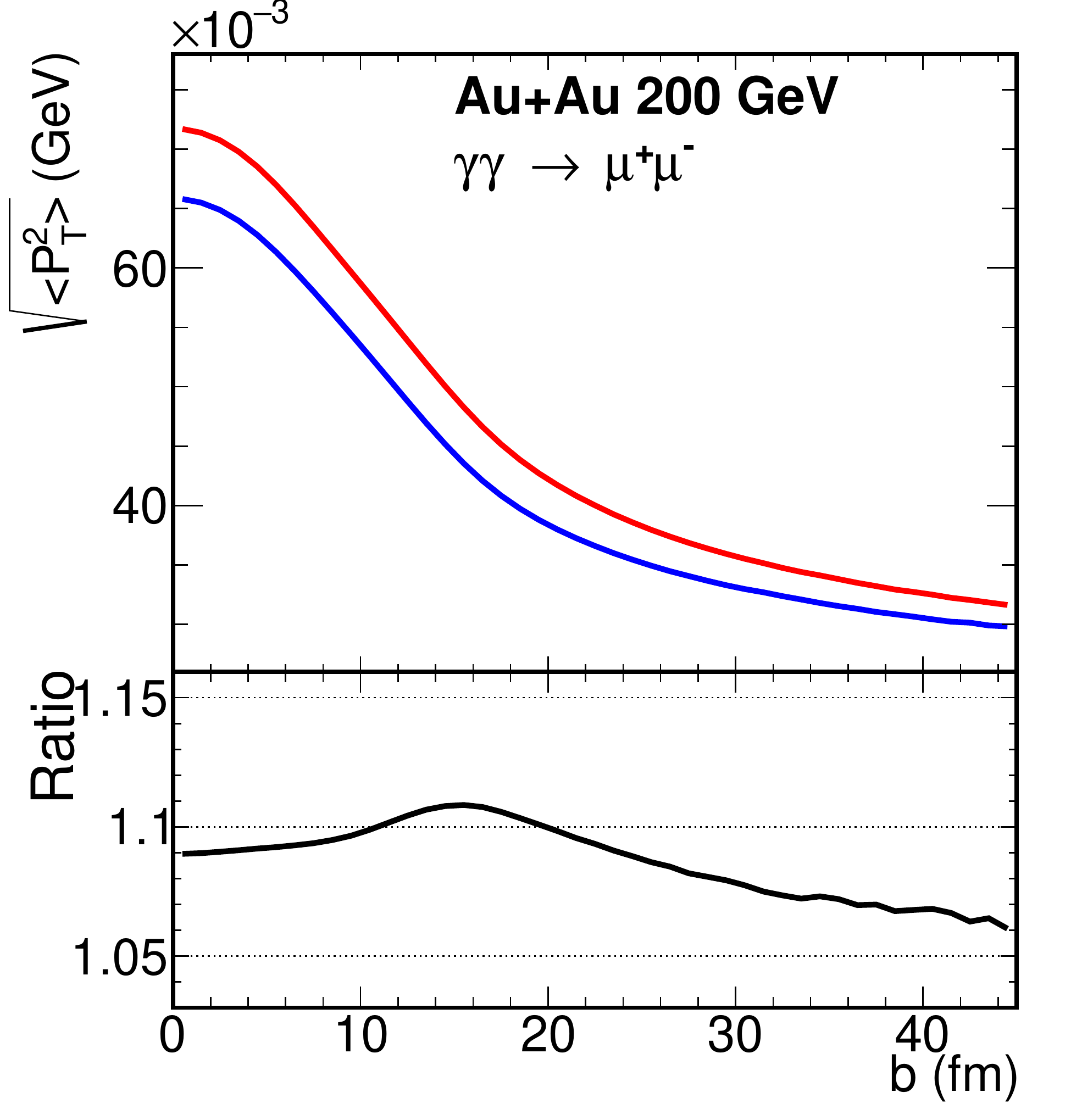}}
  \quad
  \subfloat[$\gamma \gamma \rightarrow e^+ e^-$ in Pb+Pb $\sqrt{s_{NN}}=5.02 $ TeV]{\includegraphics[width=0.33\linewidth]{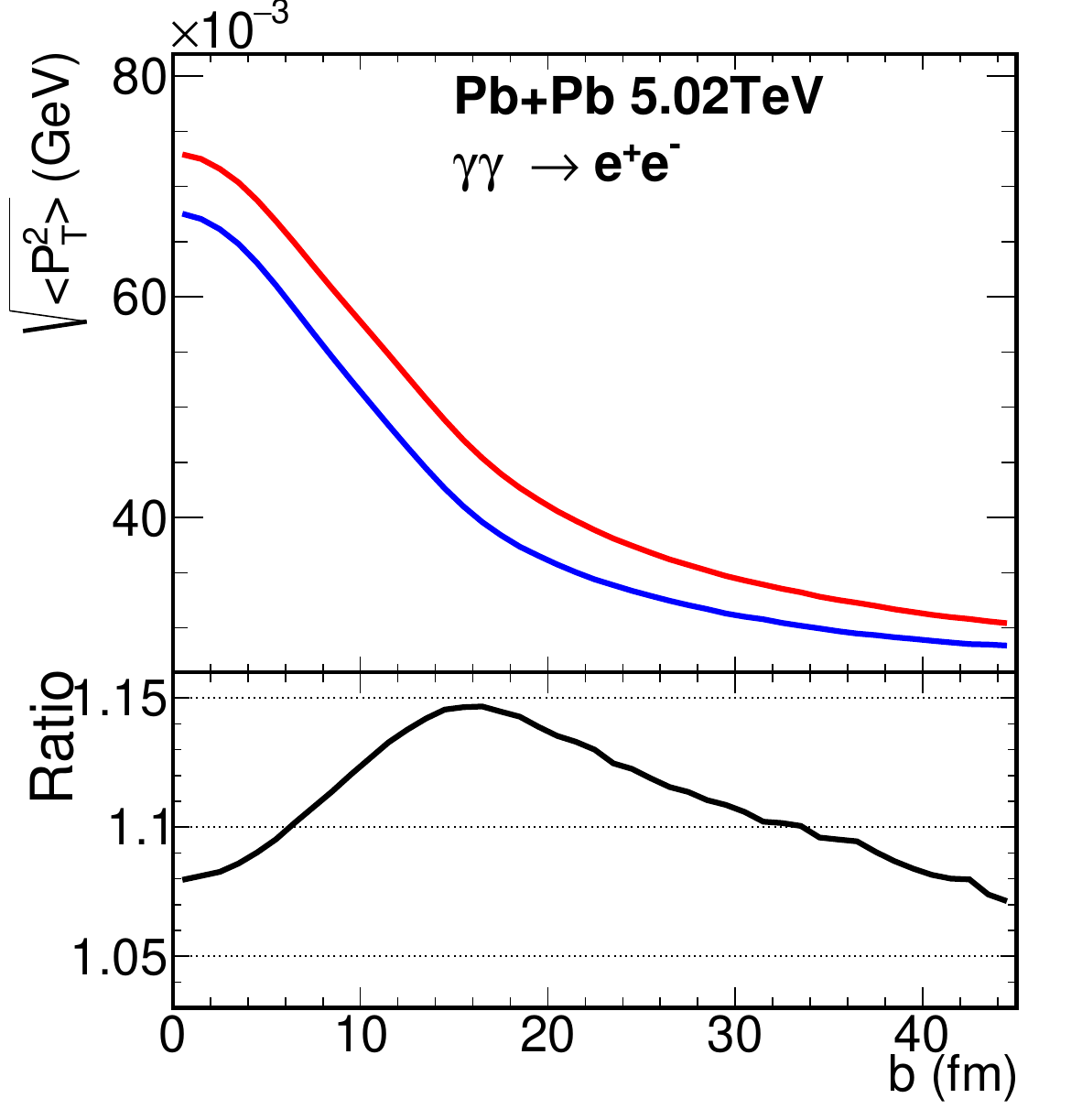}}
  \subfloat[$\gamma \gamma \rightarrow \mu^+ \mu^-$ in Pb+Pb $\sqrt{s_{NN}}=5.02 $ TeV]{\includegraphics[width=0.33\linewidth]{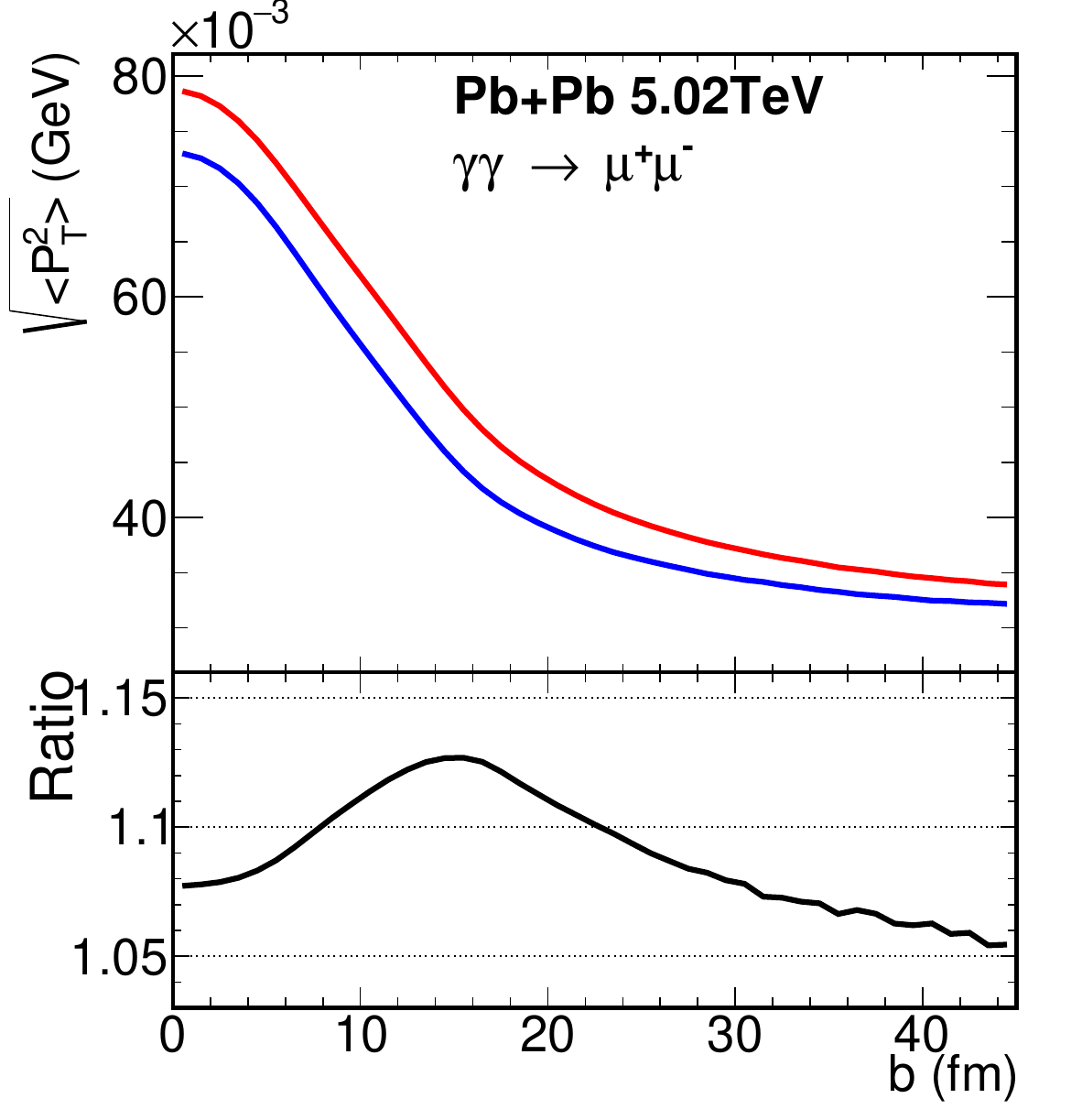}}
  \subfloat[$\gamma \gamma \rightarrow \tau^+ \tau^-$ in Pb+Pb $\sqrt{s_{NN}}=5.02 $ TeV]{\includegraphics[width=0.33\linewidth]{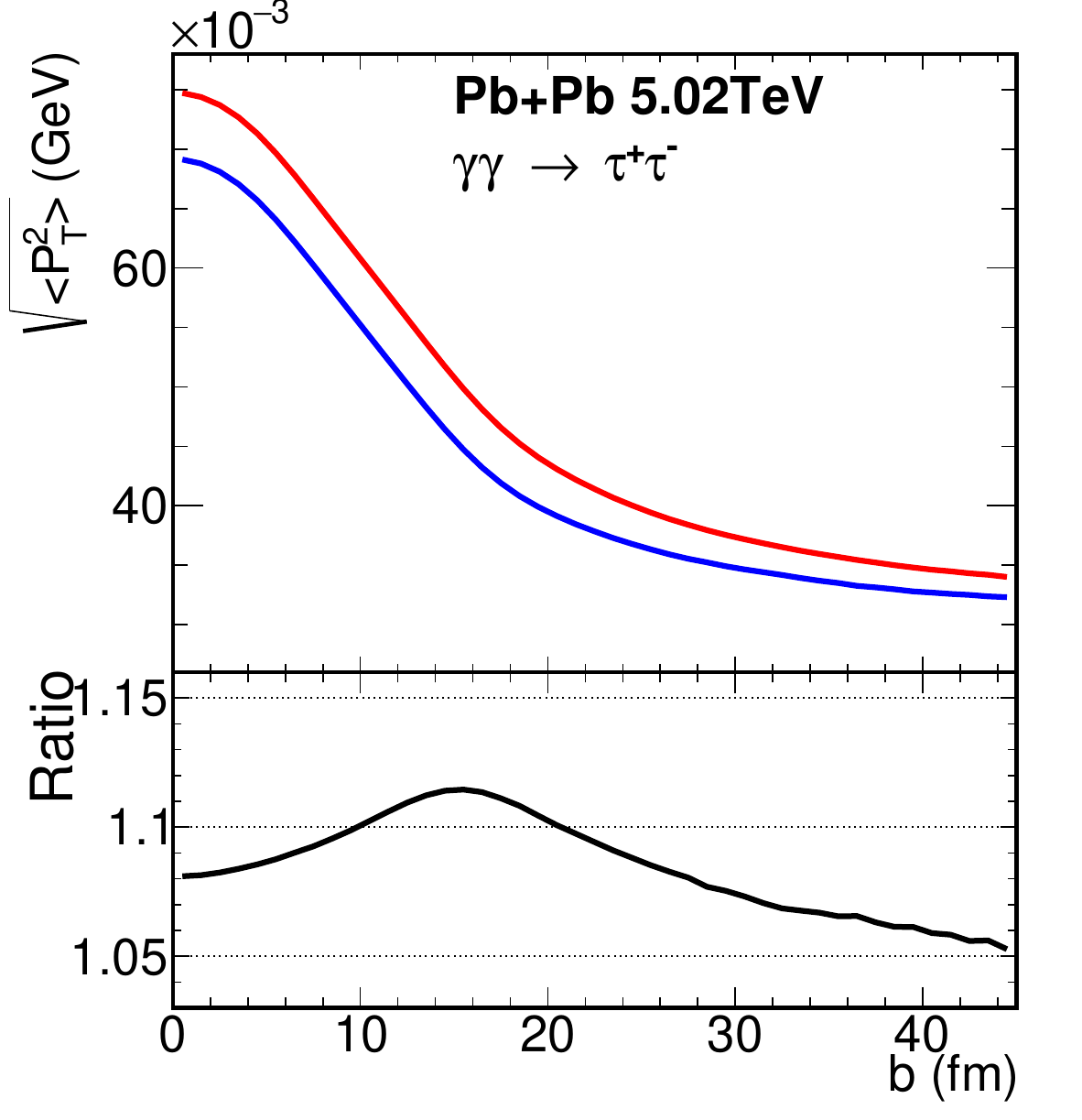}}
 \hfill
\caption{$\sqrt{\langle P_{T}^2 \rangle }$ as a function of impact parameter from the lowest-order and higher-order calculations and the ratios of higher-order (red lines) results to the lowest-order (blue lines) results at typical RHIC and LHC collision energies for different beam species. The fiducial cuts for calculations are described in Tab.~\ref{table}.}
\label{t}
\end{figure*}
\begin{figure*}[ht]
  \centering
  \subfloat[$A_{4 \phi}$ for $\gamma \gamma \rightarrow e^+ e^-$ in Au+Au 200 GeV]{\includegraphics[width=0.33\linewidth]{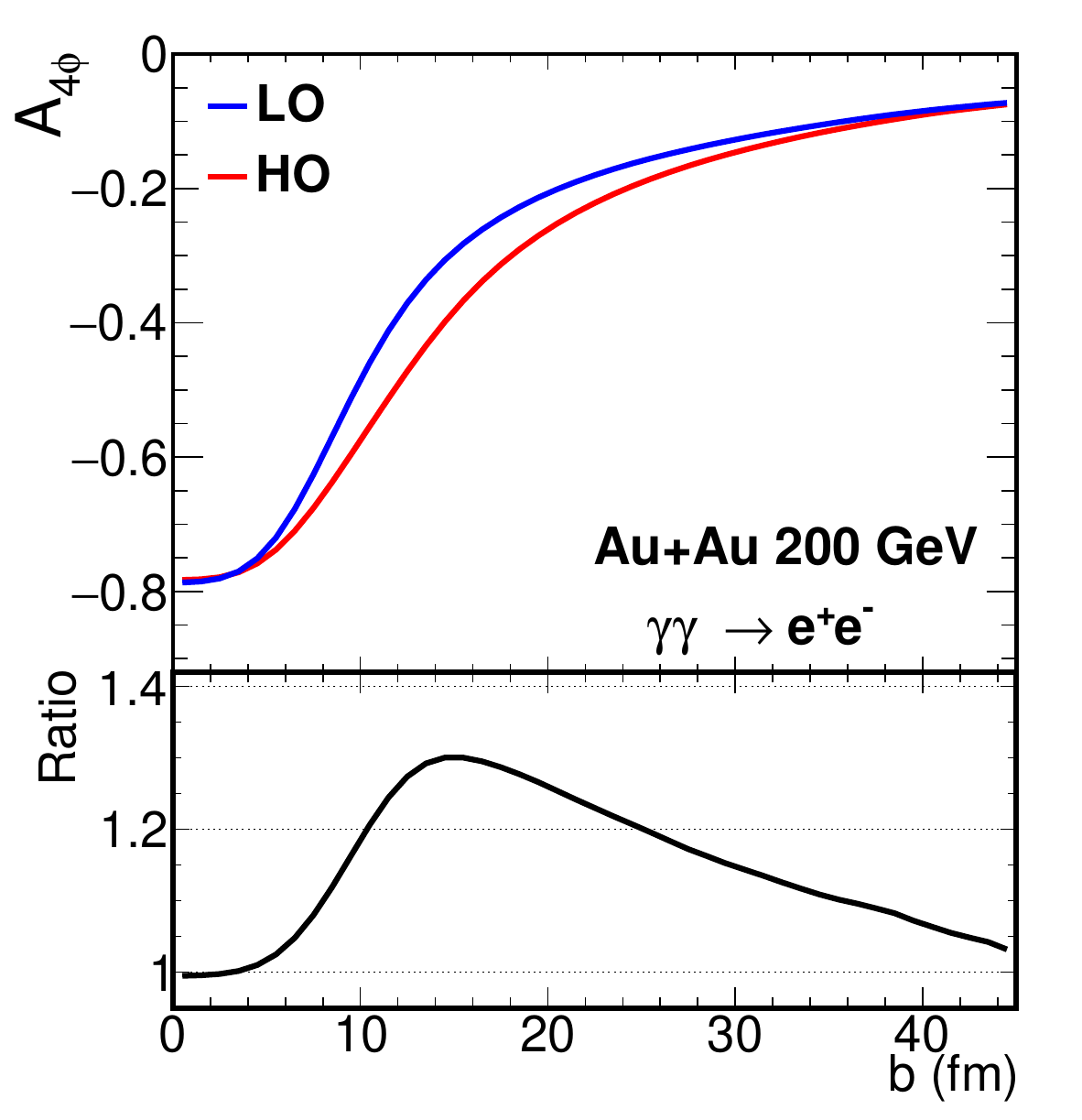}}
  \subfloat[$A_{4 \phi}$ for $\gamma \gamma \rightarrow \mu^+ \mu^-$ in Au+Au 200 GeV]{\includegraphics[width=0.33\linewidth]{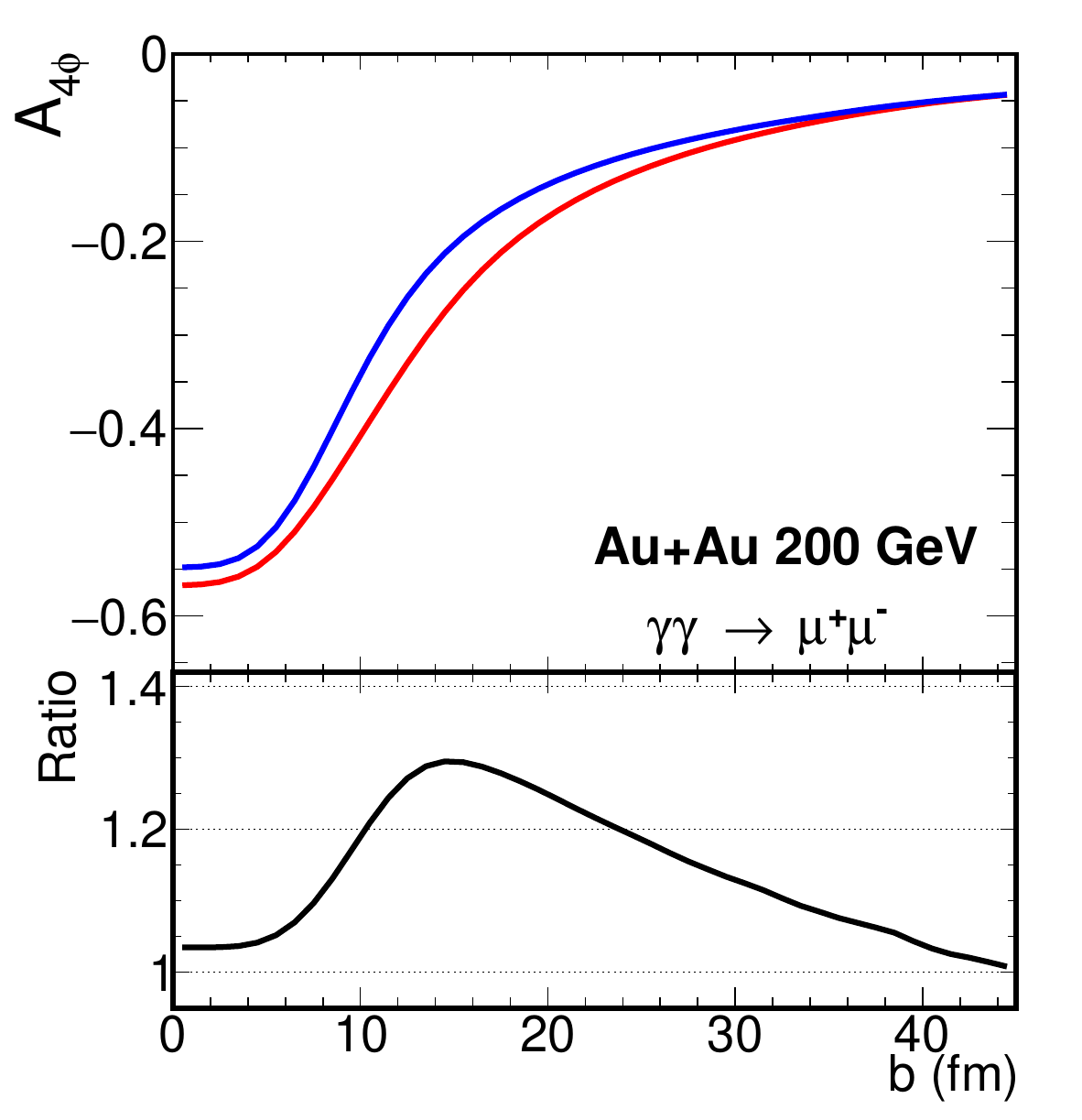}}
  \subfloat[$A_{2 \phi}$ for $\gamma \gamma \rightarrow \mu^+ \mu^-$ in Au+Au 200 GeV]{\includegraphics[width=0.33\linewidth]{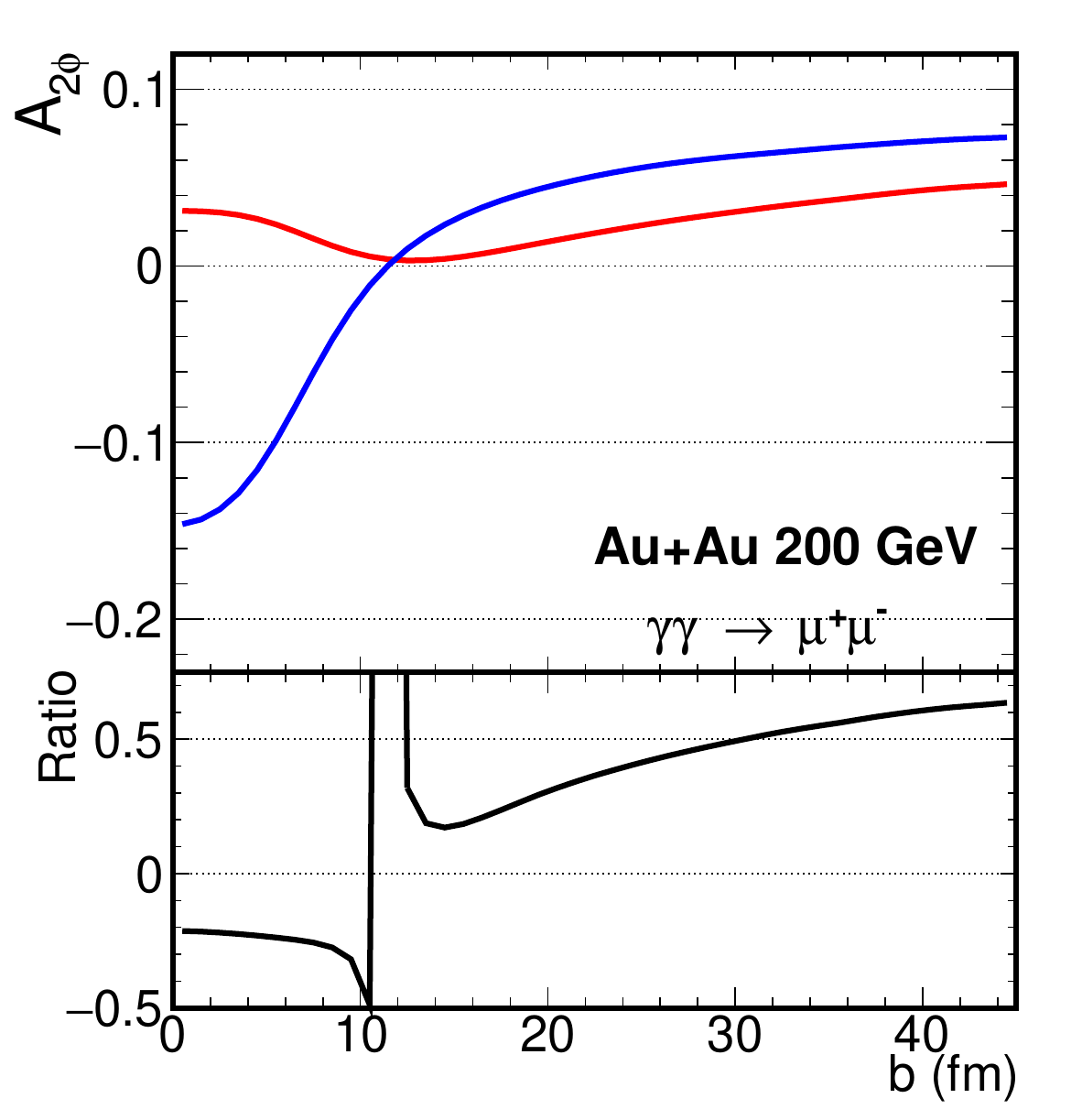}}
  \quad
  \subfloat[$A_{4 \phi}$ for $\gamma \gamma \rightarrow e^+ e^-$ in Pb+Pb 5.02 TeV]{\includegraphics[width=0.33\linewidth]{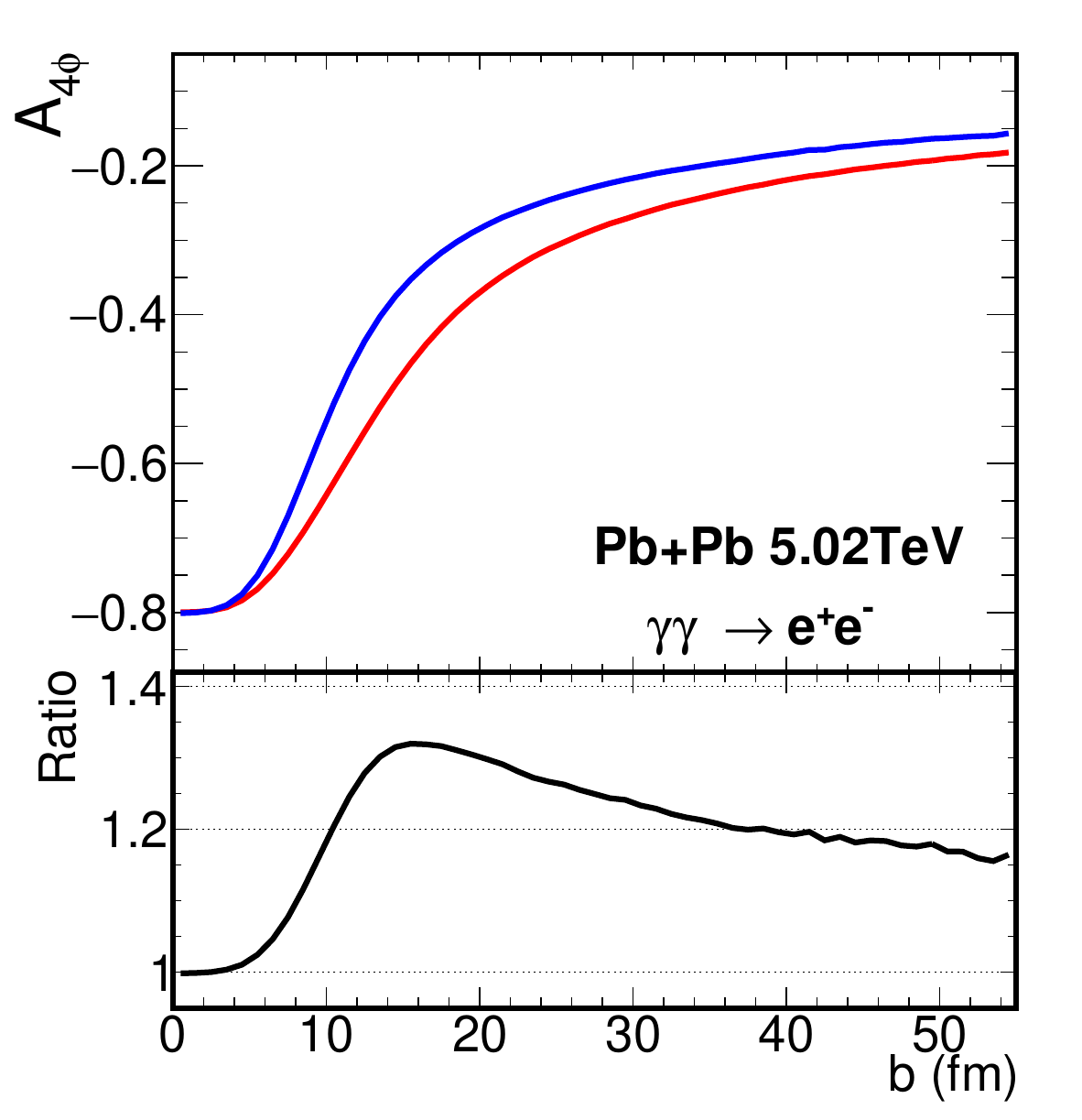}}
  \subfloat[$A_{4 \phi}$ for $\gamma \gamma \rightarrow \mu^+ \mu^-$ in Pb+Pb 5.02 TeV]{\includegraphics[width=0.33\linewidth]{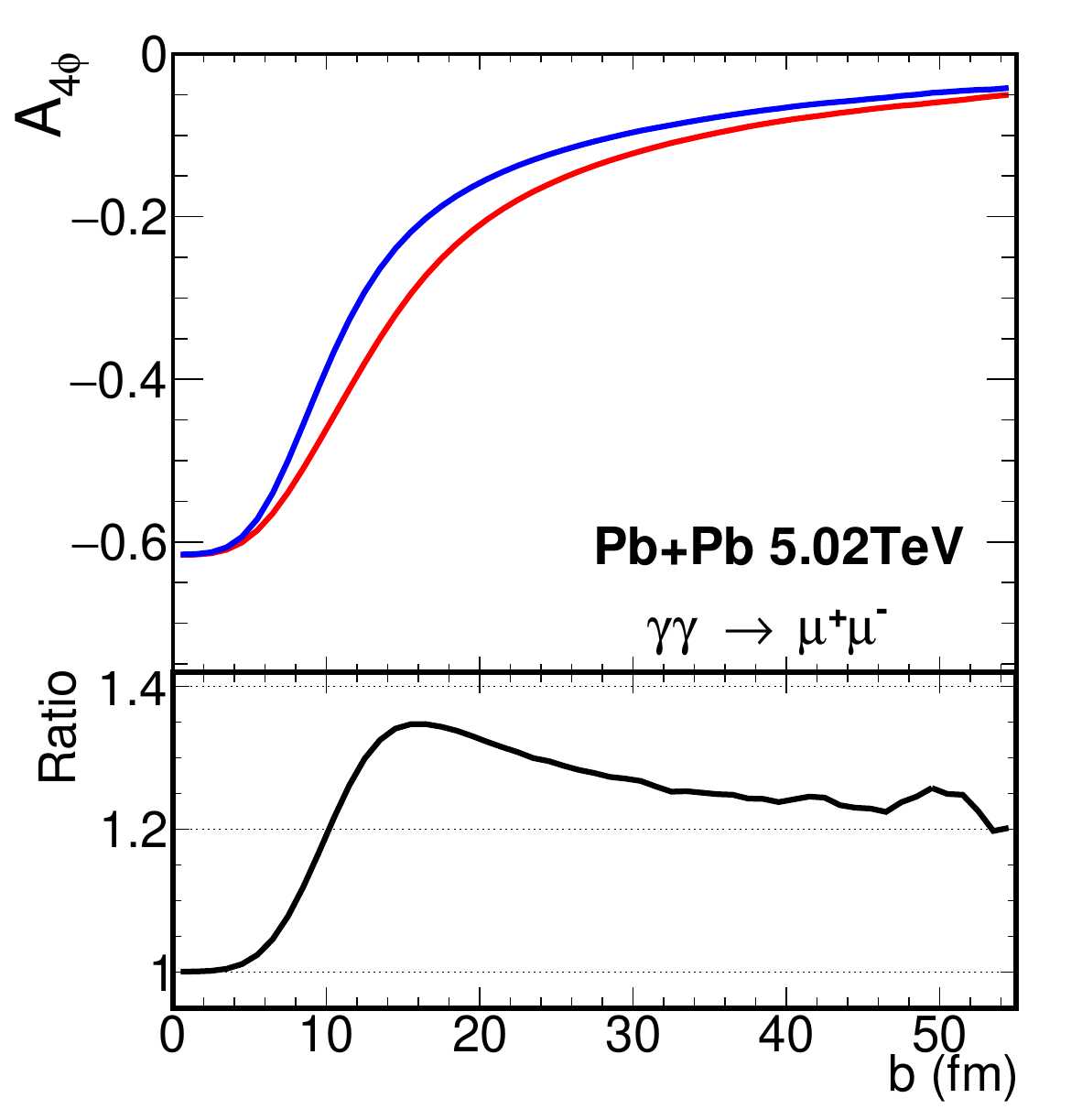}}
  \subfloat[$A_{2 \phi}$ for $\gamma \gamma \rightarrow \tau^+ \tau^-$ in Pb+Pb 5.02 TeV]{\includegraphics[width=0.33\linewidth]{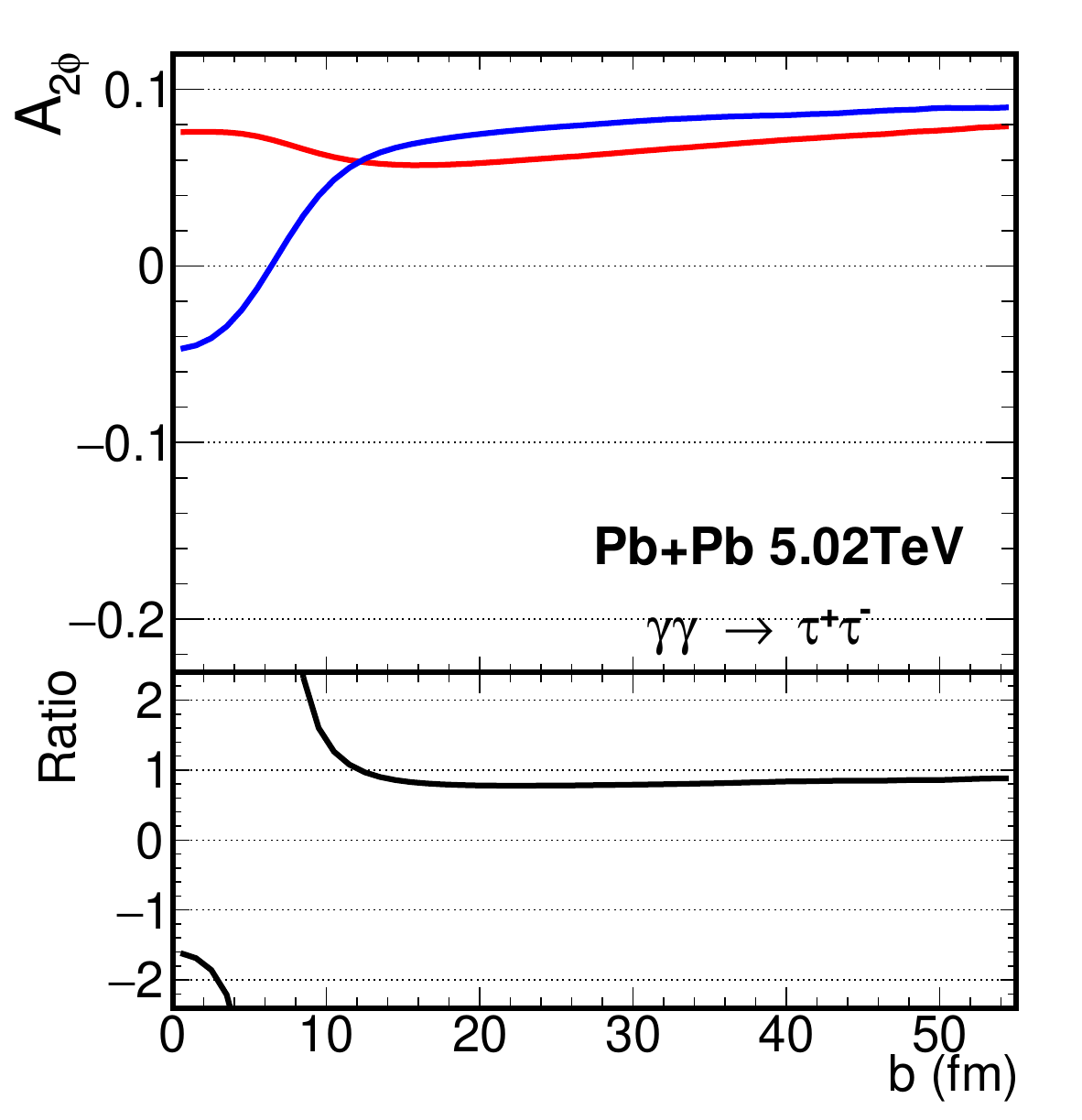}}
\caption{Amplitudes of 2nd ($A_{2 \phi}$) and 4th ($A_{4 \phi}$) modulation as a function of impact parameter from the lowest-order and higher-order calculations and the ratios of higher-order results (red lines) to the lowest-order (blue lines) results at typical RHIC and LHC collision energies for different beam species. The fiducial cuts for calculations are described in Tab.~\ref{table}.}
 \label{Aphi}
\end{figure*}

%% file: Theoretical1.tex
\section{Theoretical framework}
\label{theo}

Following the procedure of the external classical field approach outlined in Ref.~\cite{Vidovic:1992ik}, the electromagnetic potentials of the fast-moving nuclei in the Lorentz gauge can be expressed as:
\begin{equation}
A_\mu^{(1,2)}(q)=-2 \pi Z e u_\mu^{(1,2)} \delta\left(q u^{(1,2)}\right) \frac{f\left(q^2\right)}{q^2} \exp (iqb / 2).
\end{equation}
Here, $u^{(1,2)}$ represents the four-velocity of ions 1 and 2 with electric charge Z, and b denotes the impact parameter. The form factor $f(q^2)$ is the Fourier transform of the nuclear charge distribution. The lowest-order two-photon interaction for lepton pair creation involves two contributing Feynman diagrams, as illustrated in Fig.~\ref{feynman}.

With the direct and cross Feynman diagrams, the matrix element can be expressed as~\cite{Hencken:1994my}:

\begin{equation}
M=\bar{u}\left(p_{-}\right) \hat{M} v\left(p_{+}\right).    
\end{equation}

with
\begin{equation}
\begin{aligned}
\hat{M}= & -i e^2 \int \frac{d^4 q_1}{(2 \pi)^4} \slashed{A}^{(1)}\left(q_1\right) \frac{\slashed{p}_{-}-\slashed{q}_{1}+m}{\left(p_{-}-q_1\right)^2-m^2} \slashed{A}^{(2)}\left(q_2\right) \\
& -i e^2 \int \frac{d^4 q_1}{(2 \pi)^4} \slashed{A}^{(2)}\left(q_2\right) \frac{\slashed{q}_{1}-\slashed{p}_{+}+m}{\left(q_1-p_{+}\right)^2-m^2} \slashed{A}^{(1)}\left(q_1\right) \\
= & -i\left(\frac{Z e^2}{2 \pi}\right)^2 \frac{1}{2 \beta} \int d^2 q_{1 \perp} \frac{1}{q_1^2} \frac{1}{\left(q_2\right)^2} \exp \left(\mathrm{i} q_{1 \perp} \mathrm{b}\right).
\end{aligned}
\end{equation}
Where $q_2 \equiv p_{+}+p_{-}-q_1$. The probability of the lowest order pair creation can then be straightforwardly obtained as:

\begin{equation}
\begin{aligned}
& P\left(p_{+}, p_{-}, b\right) = \sum_s|M|^2 \\
&=\frac{4}{\beta^2} \int d^2 q_{1 \perp} d^2 \Delta q_{\perp} \exp \left(\mathrm{i \Delta q}_{\perp} \mathrm{b}\right) \prod_{i=0,1,3,4} f\left(N_i\right) F\left(N_i\right) \\
& \times \operatorname{Tr}\left\{\left(\slashed{p}_{-}+m\right) \left[N_{2 D}^{-1} \slashed{\omega}_1\left(\slashed{p}_{-}-\slashed{q}_1+m\right) \slashed{\omega}_2 \right.\right.\\
&\left. \hspace{5em}+ N_{2 X}^{-1} \slashed{\omega}_2\left(\slashed{q}_1-\slashed{ p}_{+}+m\right)\slashed{\omega}_1 \right]\\
& \hspace{2em} \left(\slashed{p}_{+}-m\right)\left[N_{5 D}^{-1} \slashed{\omega}_2\left(\slashed{p}_{-}-\slashed{q}_1-\slashed{q}+m\right) \slashed{\omega}_1 \right.\\
&\left.\left. \hspace{5em} +N_{5 X}^{-1} \slashed{\omega}_1\left(\slashed{q}_1+\slashed{q}-\slashed{p}_{+}+m\right)\slashed{\omega}_2\right]\right\}.
\end{aligned}
\end{equation}

In the above equations, the integration variable $\Delta q_{\perp}$ is defined as the transverse projection of $\Delta q \equiv q_{1} -q_{1}^{\prime}$. The various quantities $N_0$, $N_1$, $N_3$, $N_4$, $N_{2 D}$, $N_{2 X}$, $N_{5 D}$, and $N_{5 X}$ are defined as follows:
\begin{equation}
\begin{aligned}
&N_0 =-q_1^2, \quad N_1=-\left[q_1-\left(p_{+}+p_{-}\right)\right]^2, \\
&N_3  =-\left(q_1+\Delta q\right)^2,\quad N_4=-\left[\Delta q+\left(q_1-p_{+}-p_{-}\right)\right]^2, \\
&N_{2 D} =-\left(q_1-p_{-}\right)^2+m^2, \\
&N_{2 X} =-\left(q_1-p_{+}\right)^2+m^2, \\
&N_{5 D} =-\left(q_1+\Delta q-p_{-}\right)^2+m^2, \\
&N_{5 X} =-\left(q_1+\Delta q-p_{+}\right)^2+m^2.
\end{aligned}    
\end{equation}
The photon propagator $F(N_k)$ describes the interaction with the Coulomb field of one of the nuclei, and its regularization is crucial for including higher-order effects. To achieve this, Lee and Milstein introduced a screening of the Coulomb potential  $V(\rho,z)$  in the photon propagator~\cite{PhysRevA.61.032103, PhysRevA.64.032106}, which naturally incorporates higher-order corrections. The expression for the photon propagator is then given by:
\begin{equation}
F(\mathbf{k})=  2 \pi \int d \rho \rho J_0(k \rho) \left\{\exp \left[2 i Z \alpha K_0\left(\rho \omega / \gamma\right)\right]-1\right\}.
\label{HOFF}
\end{equation} 
In the limit as $ Z\alpha \rightarrow 0$, the perturbation theory form can be retrieved from Eq.~\ref{HOFF}, yielding
$F^0(k)=4 \pi i Z \alpha /(k^2+\omega^2 / \gamma^2)$.

%% file: results.tex
\section{Results}
\label{observable}

The experimental measurements of the Breit-Wheeler process are limited by the detector acceptance and are conducted within a fiducial phase space defined by a minimum transverse momentum cutoff and a specific pseudo-rapidity region. Thus, we calculate the differential observables for the Breit-Wheeler process using acceptance criteria that align with the typical requirements of RHIC and LHC experiments. Detailed fiducial cuts used in the calculation are provided in Tab.~\ref{table}.

Our previous work~\cite{zha2021discovery} only presented and compared the total cross-section with higher-order corrections to various experimental results. It is of great interest to differentially study the impact parameter dependence of the higher-order effect on cross-section. Figure~\ref{yield} presents the cross-section with fiducial cuts as a function of impact parameters obtained from both lowest-order and higher-order calculations. It also displays the ratios of the higher-order results to the lowest-order results at typical RHIC and LHC collision energies for different beam species. The higher-order correction leads to a reduction in the cross-section, with the reduction factor becoming more pronounced towards central collisions characterized by a small impact parameter. This behavior is expected due to the intensification of the electromagnetic field towards small impact parameters, leading to a more substantial higher-order correction~\cite{PhysRevC.59.841}. Furthermore, the higher-order QED effect shows no significant dependence on the lepton species involved in the Breit-Wheeler process for relativistic heavy-ion collisions.

In relativistic heavy-ion collisions, lepton pairs generated by QED fields exhibit a distinctive characteristic of being nearly back-to-back in azimuth and possessing small total transverse momenta ($P_\perp ~\sim \omega/\gamma$). This feature makes the Breit-Wheeler process a promising probe for investigating the electromagnetic properties of the hot nuclear medium formed during collisions with nuclear overlap. The STAR collaboration at RHIC and the ATLAS collaboration at the LHC have observed a substantial broadening effect in the transverse momentum distribution of lepton pairs resulting from photon-photon collisions in hadronic heavy-ion collisions (HHICs)~\cite{adam2018low, aaboud2018observation}, compared to those in ultra-peripheral collisions (UPCs) and calculations based on EPA. This broadening effect has been attributed to either the presence of a trapped magnetic field or QED multiple scattering within the hot medium. In our previous work~\cite{Zha:2018tlq, Zha:2017jch, Zha:2018ywo}, we initially proposed that the broadening primarily originates from the initial strength of the electromagnetic field, which varies significantly with the impact parameter. Through lowest-order QED calculations, we demonstrated a strong impact parameter dependence of the baseline $P_\perp$ broadening, leaving minimal room for additional broadening caused by the hot medium. This impact parameter dependence of the broadening baseline has been further validated by CMS measurements in UPCs~\cite{CMS:2020avp}. In the following, we will show the influence of higher-order effects on the broadening baseline and explore whether the transverse momentum broadening can serve as an indicator of the presence of higher-order QED effects.

Here, the transverse momentum broadening is characterized by the root mean square of the average transverse momentum squared, denoted as $\sqrt{\langle P_{T}^2 \rangle}$. Figure~\ref{t} illustrates $\sqrt{\langle P_{T}^2 \rangle}$ as a function of the impact parameter, obtained from both lowest-order and higher-order calculations, along with the ratios of higher-order results to the lowest-order results for typical RHIC and LHC collision energies and beam species. The inclusion of higher-order corrections enhances the transverse momentum broadening, thereby reducing the available space for additional effects originating from the hot medium in collisions with nuclear overlap. The enhancement factor of $\sqrt{\langle P_{T}^2 \rangle}$ resulting from higher-order corrections exhibits a gradual increase followed by a decrease as the impact parameter increases, revealing a prominent peak structure near 15 fm. This peak location remains unaffected by collision energy and beam species. Although the enhancement factor from higher-order corrections is approximately 10$\%$, significantly smaller than the reduction factor observed for the cross section (20-30$\%$), it does not imply that the transverse momentum broadening is less effective in detecting the presence of higher-order QED effects compared to cross-section measurements. The advantage lies in the fact that, for transverse momentum broadening measurements, uncertainties arising from luminosity determination can be fully cancelled out, and systematic uncertainties can be greatly mitigated. Both RHIC and the LHC have planned future data collections that will substantially reduce statistical uncertainties. However, the uncertainties stemming from luminosity determination and systematic effects will remain largely immune to statistical accumulation. Consequently, the $\sqrt{\langle P_{T}^2 \rangle}$ observable is more sensitive than the cross section when considering a large data sample.

Another distinctive feature of the Breit-Wheeler process in heavy-ion collisions is the complete linear polarization of the quasi-real photons produced by the strongly contracted electromagnetic field of the colliding nuclei. Li et al. proposed that this linear polarization induces second-order and fourth-order modulations in the azimuthal angle $\phi$ (in the plane perpendicular to the beam direction)~\cite{Li:2019sin}. The polarization angle $\phi$ is defined as the angle between the transverse momentum vector $\vec{P}_{T} \equiv \vec{p}_{1T} + \vec{p}_{2T}$ and the relative transverse momentum vector $\Delta \vec{P}_T \equiv \vec{p}_{1T} - \vec{p}_{2T}$, where $\vec{p}_{1T}$ and $\vec{p}_{2T}$ are the transverse momenta of the final state leptons, respectively. The strengths of the second-order and fourth-order modulations denoted as $A_{2 \phi}$ and $A_{4 \phi}$, can be extracted using the following expressions:
\begin{equation}
A_{2 \phi} = \frac{2\int d\phi \cos 2\phi f(\phi)}{\int d\phi f(\phi)}, \quad A_{4 \phi} = \frac{2\int d\phi \cos 4\phi f(\phi)}{\int d\phi f(\phi)},
\end{equation}
where $f(\phi)$ represents the differential cross section as a function of $\phi$. The angular modulation behavior has been experimentally confirmed by the STAR Collaboration in the production of dielectron pairs through the Breit-Wheeler process~\cite{STAR:2019wlg}. The experimental results can be reasonably described by lowest-order QED calculations; however, there are significant uncertainties associated with the measurements. The future data collection plans at RHIC and LHC offer the potential to significantly reduce the experimental uncertainties for angular modulation, thus providing an opportunity to unveil higher-order effects that may be hidden within the current measurement uncertainties. 

Figure~\ref{Aphi} shows the amplitudes of the second-order and fourth-order modulations as a function of impact parameters. The plot includes results from both lowest-order and higher-order calculations, as well as the ratios of the higher-order results to the lowest-order results, for typical beam species at RHIC and LHC collision energies. Regarding the fourth-order modulation, the higher-order correction amplifies its magnitude. The modification factor resulting from the higher-order effect exhibits a gradual increase followed by a decrease as the impact parameter increases, revealing a prominent peak structure near 15 fm. This behavior closely resembles the pattern observed for $\sqrt{\langle P_{T}^2 \rangle}$. In the very small ($b \rightarrow 0$ fm) or large ($b \rightarrow \infty$ fm) impact parameter regions, the modification factors approach unity. This suggests that the angular modulation observable is only sensitive to higher-order effects in the moderate impact parameter region. For the second-order modulation, a distinct sign transition is observed in the lowest-order calculation, but it is absent in the higher-order calculations. Therefore, the ratios of higher-order to lowest-order results can serve as a powerful tool to determine the presence or absence of higher-order QED effects. The angular modulation observables are self-normalized, which shares the same advantage as $\sqrt{\langle P_{T}^2 \rangle}$ in terms of cancelling out the effects of luminosity determination and mitigating systematic uncertainties.

The angular modulation strength in the Breit-Wheeler process is associated with the lepton mass, denoted as $m$, and the transverse momentum difference $\Delta P_{T}$, as derived in Ref.~\cite{Li:2019yzy}. The expressions for $A_{2\phi}$ and $A_{4\phi}$ can be approximated as follows:

\begin{equation}
\begin{aligned}
A_{2\phi} &\propto \frac{4 m^2 \Delta P_{T}^2}{\left(m^2+\Delta P_{T}^2\right)^2} \sim \frac{4 m^2}{\Delta P_{T}^2} \quad \text{for } \Delta P_{T}^2 \gg m, \\
A_{4\phi} &\propto \frac{- 2 \Delta P_{T}^4} {\left(m^2+\Delta P_{T}^2\right)^2} \sim -2\left(1-\frac{m^2}{\Delta P_{T}^2}\right) \quad \text{for } \Delta P_{T}^2 \gg m.
\end{aligned}
\end{equation}

Based on this analytical relationship, it is inferred that $A_{2\phi}$ becomes negligible in the large $\Delta P_{T}$ regime. Therefore, for electron pair production in RHIC and muon pair production in LHC, the $A_{2\phi}$ values are extremely small and cannot be precisely measured. Consequently, these results are not presented in the paper.

%% file: Summary1.tex
\section{Summary}
Considering the significant effective QED coupling in relativistic heavy-ion collisions, the involvement of higher-order QED effects is necessary for an accurate description of the Breit-Wheeler process. Although our previous work discovered evidence of higher-order corrections in the total cross-section, the presence of substantial experimental and theoretical uncertainties prevents unambiguous verification. This study investigates the impact of higher-order QED effects on the differential distributions of the Breit-Wheeler process in relativistic heavy-ion collisions at typical RHIC and LHC collision energies and species. The higher-order correction on the differential cross-section increases in significance as the impact parameter decreases. The effect on transverse momentum broadening and the fourth-order modulation exhibits a distinct pattern characterized by a gradual increase, followed by a decrease as the impact parameter increases. Notably, it demonstrates a prominent peak structure near 15 fm. Additionally, the higher-order correction alters the sign of the second-order modulation within a specific range of impact parameters. This characteristic serves as a powerful indicator for discerning the existence or absence of higher-order QED effects. The transverse momentum broadening and angular modulation observables possess the advantage of being self-normalized, effectively cancelling out the effects of luminosity determination and mitigating systematic uncertainties. Consequently, these two observables surpass the cross-section in terms of potency when analyzing a large dataset. RHIC and LHC have scheduled a series of future production activities involving Au+Au and Pb+Pb collisions. This work has the potential to delineate a pathway for unveiling latent higher-order effects through the utilization of more substantial statistics.

%% file: acknowledge.tex
\section*{Acknowledgement}

This work is supported in part by the National Key Research and Development Program of China under Contract No. 2022YFA1604900 the National Natural Science Foundation of China (NSFC) under Contract No. 12175223 and 12005220. W. Zha is supported by Anhui Provincial Natural Science Foundation No. 2208085J23 and Youth Innovation Promotion Association of Chinese Academy of Sciences.

%% file: main.bbl
\begin{thebibliography}{10}
\expandafter\ifx\csname url\endcsname\relax
  \def\url#1{\texttt{#1}}\fi
\expandafter\ifx\csname urlprefix\endcsname\relax\def\urlprefix{URL }\fi
\expandafter\ifx\csname href\endcsname\relax
  \def\href#1#2{#2} \def\path#1{#1}\fi

\bibitem{PhysRev.46.1087}
G.~Breit, J.~A. Wheeler, Collision of two light quanta, Phys. Rev. 46 (1934)
  1087--1091.
\newblock \href {https://doi.org/10.1103/PhysRev.46.1087}
  {\path{doi:10.1103/PhysRev.46.1087}}.

\bibitem{WW}
C.~F.~v. Weizs$\ddot{a}$cker, {Ausstrahlung bei St$\ddot{o}\beta$en sehr
  schneller Elektronen}, Z. Physik 88 (1934) 612–625.
\newblock \href {https://doi.org/https://doi.org/10.1007/BF01333110}
  {\path{doi:https://doi.org/10.1007/BF01333110}}.

\bibitem{PhysRev.45.729}
E.~J. Williams, {Nature of the High Energy Particles of Penetrating Radiation
  and Status of Ionization and Radiation Formulae}, Phys. Rev. 45 (1934)
  729--730.
\newblock \href {https://doi.org/10.1103/PhysRev.45.729}
  {\path{doi:10.1103/PhysRev.45.729}}.

\bibitem{STAR:2019wlg}
J.~Adam, et~al., {Measurement of $e^+e^-$ Momentum and Angular Distributions
  from Linearly Polarized Photon Collisions}, Phys. Rev. Lett. 127~(5) (2021)
  052302.
\newblock \href {http://arxiv.org/abs/1910.12400} {\path{arXiv:1910.12400}},
  \href {https://doi.org/10.1103/PhysRevLett.127.052302}
  {\path{doi:10.1103/PhysRevLett.127.052302}}.

\bibitem{PhysRev.93.768}
H.~A. Bethe, L.~C. Maximon, Theory of bremsstrahlung and pair production. i.
  differential cross section, Phys. Rev. 93 (1954) 768--784.
\newblock \href {https://doi.org/10.1103/PhysRev.93.768}
  {\path{doi:10.1103/PhysRev.93.768}}.

\bibitem{PhysRev.93.788}
H.~Davies, H.~A. Bethe, L.~C. Maximon, Theory of bremsstrahlung and pair
  production. ii. integral cross section for pair production, Phys. Rev. 93
  (1954) 788--795.
\newblock \href {https://doi.org/10.1103/PhysRev.93.788}
  {\path{doi:10.1103/PhysRev.93.788}}.

\bibitem{BH}
H.~Bethe, W.~Heitler, On the stopping of fast particles and on the creation of
  positive electrons, Proc. R. Soc. Lond. A 146 (1934) 83--112.

\bibitem{PhysRevD.57.4025}
D.~Ivanov, K.~Melnikov, Lepton pair production by a high energy photon in a
  strong electromagnetic field, Phys. Rev. D 57 (1998) 4025--4034.
\newblock \href {https://doi.org/10.1103/PhysRevD.57.4025}
  {\path{doi:10.1103/PhysRevD.57.4025}}.

\bibitem{kk}
S.~R. Klein, {{$e^+ e^-$} pair production from 10-GeV to 10-ZeV}, Radiat. Phys.
  Chem. 75 (2006) 696--711.
\newblock \href {http://arxiv.org/abs/hep-ex/0402028}
  {\path{arXiv:hep-ex/0402028}}, \href
  {https://doi.org/10.1016/j.radphyschem.2005.09.005}
  {\path{doi:10.1016/j.radphyschem.2005.09.005}}.

\bibitem{RHIC1}
L.~Esnault, E.~d'Humi\`eres, A.~Arefiev, et~al., {Electron-positron pair
  production in the collision of real photon beams with wide energy
  distributions}, Plasma Phys. Control. Fusion 63~(12) (2021) 125015.
\newblock \href {http://arxiv.org/abs/2103.09099} {\path{arXiv:2103.09099}},
  \href {https://doi.org/10.1088/1361-6587/ac2e3e}
  {\path{doi:10.1088/1361-6587/ac2e3e}}.

\bibitem{RHIC2}
S.~Afanasiev, et~al., {Photoproduction of {J/$\psi$} and of high mass
  {$e^+e^-$} in ultra-peripheral Au+Au collisions at {$\sqrt{s}$ = 200 GeV}},
  Phys. Lett. B 679 (2009) 321--329.
\newblock \href {http://arxiv.org/abs/0903.2041} {\path{arXiv:0903.2041}},
  \href {https://doi.org/10.1016/j.physletb.2009.07.061}
  {\path{doi:10.1016/j.physletb.2009.07.061}}.

\bibitem{baltz2008physics}
A.~Baltz, G.~Baur, d’Enterria, et~al., The physics of ultraperipheral
  collisions at the lhc, Phys. Rev. 458~(1-3) (2008) 1--171.

\bibitem{PhysRevC.97.054903}
S.~R. Klein, Two-photon production of dilepton pairs in peripheral heavy ion
  collisions, Phys. Rev. C 97 (2018) 054903.
\newblock \href {https://doi.org/10.1103/PhysRevC.97.054903}
  {\path{doi:10.1103/PhysRevC.97.054903}}.

\bibitem{aaboud2018observation}
M.~Aaboud, G.~Aad, B.~Abbott, et~al., Observation of centrality-dependent
  acoplanarity for muon pairs produced via two-photon scattering in pb+ pb
  collisions at s n n= 5.02 tev with the atlas detector, Phys. Rev. Lett.
  121~(21) (2018) 212301.

\bibitem{adam2018low}
J.~Adam, et~al., {Low-$p_{T}$ $e^+ e^-$ Pair Production in Au+Au Collisions at
  $\sqrt{s_{NN}}$ = 200 GeV and U+U Collisions at $\sqrt{s_{NN}}$ = 193 GeV at
  STAR}, Phys. Rev. Lett. 121~(13) (2018) 132301.

\bibitem{starlight}
S.~R. Klein, J.~Nystrand, J.~Seger, et~al., {STARlight: A Monte Carlo
  simulation program for ultra-peripheral collisions of relativistic ions},
  Comput. Phys. Commun. 212 (2017) 258--268.
\newblock \href {http://arxiv.org/abs/1607.03838} {\path{arXiv:1607.03838}},
  \href {https://doi.org/10.1016/j.cpc.2016.10.016}
  {\path{doi:10.1016/j.cpc.2016.10.016}}.

\bibitem{Zha:2018tlq}
W.~Zha, J.~D. Brandenburg, Z.~Tang, et~al., {Initial transverse-momentum
  broadening of Breit-Wheeler process in relativistic heavy-ion collisions},
  Phys. Lett. B 800 (2020) 135089.
\newblock \href {http://arxiv.org/abs/1812.02820} {\path{arXiv:1812.02820}},
  \href {https://doi.org/10.1016/j.physletb.2019.135089}
  {\path{doi:10.1016/j.physletb.2019.135089}}.

\bibitem{BAUR20071}
G.~Baur, K.~Hencken, D.~Trautmann, Electron–positron pair production in
  ultrarelativistic heavy ion collisions, Physics Reports 453~(1) (2007) 1--27.
\newblock \href {https://doi.org/https://doi.org/10.1016/j.physrep.2007.09.002}
  {\path{doi:https://doi.org/10.1016/j.physrep.2007.09.002}}.

\bibitem{SUN2020135679}
Z.-h. Sun, D.-x. Zheng, J.~Zhou, Y.-j. Zhou, Studying coulomb correction at eic
  and eicc, Physics Letters B 808 (2020) 135679.
\newblock \href
  {https://doi.org/https://doi.org/10.1016/j.physletb.2020.135679}
  {\path{doi:https://doi.org/10.1016/j.physletb.2020.135679}}.

\bibitem{BALTZ2001395}
A.~Baltz, F.~Gelis, L.~McLerran, A.~Peshier, Coulomb corrections to e+e−
  production in ultra-relativistic nuclear collisions, Nuclear Physics A
  695~(1) (2001) 395--429.
\newblock \href {https://doi.org/https://doi.org/10.1016/S0375-9474(01)01109-5}
  {\path{doi:https://doi.org/10.1016/S0375-9474(01)01109-5}}.

\bibitem{PhysRevA.61.032103}
R.~N. Lee, A.~I. Milstein, {Coulomb corrections to the
  ${e}^{+}{e}^{\ensuremath{-}}$ pair production in ultrarelativistic heavy-ion
  collisions}, Phys. Rev. A 61 (2000) 032103.
\newblock \href {https://doi.org/10.1103/PhysRevA.61.032103}
  {\path{doi:10.1103/PhysRevA.61.032103}}.

\bibitem{PhysRevA.64.032106}
R.~N. Lee, A.~I. Milstein, {Coulomb corrections and multiple
  ${e}^{+}{e}^{\ensuremath{-}}\ensuremath{-}\mathrm{pair}$ production in
  ultrarelativistic nuclear collisions}, Phys. Rev. A 64 (2001) 032106.
\newblock \href {https://doi.org/10.1103/PhysRevA.64.032106}
  {\path{doi:10.1103/PhysRevA.64.032106}}.

\bibitem{PhysRevLett.100.062302}
A.~J. Baltz, Evidence for higher order qed effects in
  ${e}^{+}{e}^{\ensuremath{-}}$ pair production at the bnl relativistic heavy
  ion collider, Phys. Rev. Lett. 100 (2008) 062302.
\newblock \href {https://doi.org/10.1103/PhysRevLett.100.062302}
  {\path{doi:10.1103/PhysRevLett.100.062302}}.

\bibitem{zha2021discovery}
W.~Zha, Z.~Tang, Discovery of higher-order quantum electrodynamics effect for
  the vacuum pair production, J. High Energ. Phys. 2021~(8) (2021) 1--18.

\bibitem{Vidovic:1992ik}
M.~Vidovic, M.~Greiner, C.~Best, et~al., {Impact parameter dependence of the
  electromagnetic particle production in ultrarelativistic heavy ion
  collisions}, Phys. Rev. C 47 (1993) 2308--2319.
\newblock \href {https://doi.org/10.1103/PhysRevC.47.2308}
  {\path{doi:10.1103/PhysRevC.47.2308}}.

\bibitem{Hencken:1994my}
K.~Hencken, D.~Trautmann, G.~Baur, {Impact parameter dependence of the total
  probability for the electromagnetic electron - positron pair production in
  relativistic heavy ion collisions}, Phys. Rev. A 51 (1995) 1874--1882.
\newblock \href {http://arxiv.org/abs/nucl-th/9410014}
  {\path{arXiv:nucl-th/9410014}}, \href
  {https://doi.org/10.1103/PhysRevA.51.1874}
  {\path{doi:10.1103/PhysRevA.51.1874}}.

\bibitem{PhysRevC.59.841}
K.~Hencken, D.~Trautmann, G.~Baur, Calculation of higher-order effects in
  electron-positron pair production in relativistic heavy ion collisions, Phys.
  Rev. C 59 (1999) 841--844.
\newblock \href {https://doi.org/10.1103/PhysRevC.59.841}
  {\path{doi:10.1103/PhysRevC.59.841}}.

\bibitem{Zha:2017jch}
W.~Zha, S.~R. Klein, R.~Ma, et~al., {Coherent J/$\psi$ photoproduction in
  hadronic heavy-ion collisions}, Phys. Rev. C 97~(4) (2018) 044910.
\newblock \href {http://arxiv.org/abs/1705.01460} {\path{arXiv:1705.01460}},
  \href {https://doi.org/10.1103/PhysRevC.97.044910}
  {\path{doi:10.1103/PhysRevC.97.044910}}.

\bibitem{Zha:2018ywo}
W.~Zha, L.~Ruan, Z.~Tang, Z.~Xu, S.~Yang, {Coherent lepton pair production in
  hadronic heavy ion collisions}, Phys. Lett. B 781 (2018) 182--186.
\newblock \href {http://arxiv.org/abs/1804.01813} {\path{arXiv:1804.01813}},
  \href {https://doi.org/10.1016/j.physletb.2018.04.006}
  {\path{doi:10.1016/j.physletb.2018.04.006}}.

\bibitem{CMS:2020avp}
A.~M. Sirunyan, A.~Tumasyan, W.~Adam, et~al., {Observation of forward neutron
  multiplicity dependence of dimuon acoplanarity in ultra-peripheral PbPb
  collisions at $\sqrt{s_{\mathrm{NN}}} = 5.02~\mathrm{TeV}$} (2020).

\bibitem{Li:2019sin}
C.~Li, J.~Zhou, Y.-J. Zhou, {Impact parameter dependence of the azimuthal
  asymmetry in lepton pair production in heavy ion collisions}, Phys. Rev. D
  101~(3) (2020) 034015.
\newblock \href {http://arxiv.org/abs/1911.00237} {\path{arXiv:1911.00237}},
  \href {https://doi.org/10.1103/PhysRevD.101.034015}
  {\path{doi:10.1103/PhysRevD.101.034015}}.

\bibitem{Li:2019yzy}
C.~Li, J.~Zhou, Y.-J. Zhou, {Probing the linear polarization of photons in
  ultraperipheral heavy ion collisions}, Phys. Lett. B 795 (2019) 576--580.
\newblock \href {http://arxiv.org/abs/1903.10084} {\path{arXiv:1903.10084}},
  \href {https://doi.org/10.1016/j.physletb.2019.07.005}
  {\path{doi:10.1016/j.physletb.2019.07.005}}.

\end{thebibliography}
